\documentclass[ALICE,manyauthors]{cernphprep}
\usepackage[comma,square,numbers,sort&compress]{natbib}

\usepackage{lineno}
\usepackage{xspace}
\usepackage{hyperref}
\usepackage{color}
\usepackage{ amssymb }
\usepackage{amsmath}
\usepackage[T1]{fontenc}

\begin{document}
%

\newcommand{\pp}           {pp\xspace}
\newcommand{\ppbar}        {\mbox{$\mathrm {p\overline{p}}$}\xspace}
\newcommand{\XeXe}         {\mbox{Xe--Xe}\xspace}
\newcommand{\PbPb}         {\mbox{Pb--Pb}\xspace}
\newcommand{\pA}           {\mbox{pA}\xspace}
\newcommand{\pPb}          {\mbox{p--Pb}\xspace}
\newcommand{\AuAu}         {\mbox{Au--Au}\xspace}
\newcommand{\dAu}          {\mbox{d--Au}\xspace}

\newcommand{\s}            {\ensuremath{\sqrt{s}}\xspace}
\newcommand{\snn}          {\ensuremath{\sqrt{s_{\mathrm{NN}}}}\xspace}
\newcommand{\pt}           {\ensuremath{p_{\rm T}}\xspace}
\newcommand{\meanpt}       {$\langle p_{\mathrm{T}}\rangle$\xspace}
\newcommand{\ycms}         {\ensuremath{y_{\rm CMS}}\xspace}
\newcommand{\ylab}         {\ensuremath{y_{\rm lab}}\xspace}
\newcommand{\etarange}[1]  {\mbox{$\left | \eta \right |~<~#1$}}
\newcommand{\yrange}[1]    {\mbox{$\left | y \right |~<~#1$}}
\newcommand{\dndy}         {\ensuremath{\mathrm{d}N_\mathrm{ch}/\mathrm{d}y}\xspace}
\newcommand{\dndeta}       {\ensuremath{\mathrm{d}N_\mathrm{ch}/\mathrm{d}\eta}\xspace}
\newcommand{\avdndeta}     {\ensuremath{\langle\dndeta\rangle}\xspace}
\newcommand{\dNdy}         {\ensuremath{\mathrm{d}N_\mathrm{ch}/\mathrm{d}y}\xspace}
\newcommand{\Npart}        {\ensuremath{N_\mathrm{part}}\xspace}
\newcommand{\Ncoll}        {\ensuremath{N_\mathrm{coll}}\xspace}
\newcommand{\dEdx}         {\ensuremath{\textrm{d}E/\textrm{d}x}\xspace}
\newcommand{\RpPb}         {\ensuremath{R_{\rm pPb}}\xspace}

\newcommand{\nineH}        {$\sqrt{s}~=~0.9$~Te\kern-.1emV\xspace}
\newcommand{\seven}        {$\sqrt{s}~=~7$~Te\kern-.1emV\xspace}
\newcommand{\twoH}         {$\sqrt{s}~=~0.2$~Te\kern-.1emV\xspace}
\newcommand{\twosevensix}  {$\sqrt{s}~=~2.76$~Te\kern-.1emV\xspace}
\newcommand{\five}         {$\sqrt{s}~=~5.02$~Te\kern-.1emV\xspace}
\newcommand{\twosevensixnn}{$\sqrt{s_{\mathrm{NN}}}~=~2.76$~Te\kern-.1emV\xspace}
\newcommand{\fivenn}       {$\sqrt{s_{\mathrm{NN}}}~=~5.02$~Te\kern-.1emV\xspace}
\newcommand{\LT}           {L{\'e}vy-Tsallis\xspace}
\newcommand{\GeVc}         {Ge\kern-.1emV/$c$\xspace}
\newcommand{\MeVc}         {Me\kern-.1emV/$c$\xspace}
\newcommand{\TeV}          {Te\kern-.1emV\xspace}
\newcommand{\GeV}          {Ge\kern-.1emV\xspace}
\newcommand{\MeV}          {Me\kern-.1emV\xspace}
\newcommand{\GeVmass}      {Ge\kern-.2emV/$c^2$\xspace}
\newcommand{\MeVmass}      {Me\kern-.2emV/$c^2$\xspace}
\newcommand{\lumi}         {\ensuremath{\mathcal{L}}\xspace}

\newcommand{\ITS}          {\rm{ITS}\xspace}
\newcommand{\TOF}          {\rm{TOF}\xspace}
\newcommand{\ZDC}          {\rm{ZDC}\xspace}
\newcommand{\ZDCs}         {\rm{ZDCs}\xspace}
\newcommand{\ZNA}          {\rm{ZNA}\xspace}
\newcommand{\ZNC}          {\rm{ZNC}\xspace}
\newcommand{\SPD}          {\rm{SPD}\xspace}
\newcommand{\SDD}          {\rm{SDD}\xspace}
\newcommand{\SSD}          {\rm{SSD}\xspace}
\newcommand{\TPC}          {\rm{TPC}\xspace}
\newcommand{\TRD}          {\rm{TRD}\xspace}
\newcommand{\VZERO}        {\rm{V0}\xspace}
\newcommand{\VZEROA}       {\rm{V0A}\xspace}
\newcommand{\VZEROC}       {\rm{V0C}\xspace}
\newcommand{\Vdecay} 	   {\ensuremath{V^{0}}\xspace}

\newcommand{\ee}           {\ensuremath{e^{+}e^{-}}} 
\newcommand{\pip}          {\ensuremath{\pi^{+}}\xspace}
\newcommand{\pim}          {\ensuremath{\pi^{-}}\xspace}
\newcommand{\kap}          {\ensuremath{\rm{K}^{+}}\xspace}
\newcommand{\kam}          {\ensuremath{\rm{K}^{-}}\xspace}
\newcommand{\pbar}         {\ensuremath{\rm\overline{p}}\xspace}
\newcommand{\kzero}        {\ensuremath{{\rm K}^{0}_{\rm{S}}}\xspace}
\newcommand{\lmb}          {\ensuremath{\Lambda}\xspace}
\newcommand{\almb}         {\ensuremath{\overline{\Lambda}}\xspace}
\newcommand{\Om}           {\ensuremath{\Omega^-}\xspace}
\newcommand{\Mo}           {\ensuremath{\overline{\Omega}^+}\xspace}
\newcommand{\X}            {\ensuremath{\Xi^-}\xspace}
\newcommand{\Ix}           {\ensuremath{\overline{\Xi}^+}\xspace}
\newcommand{\Xis}          {\ensuremath{\Xi^{\pm}}\xspace}
\newcommand{\Oms}          {\ensuremath{\Omega^{\pm}}\xspace}
\newcommand{\degree}       {\ensuremath{^{\rm o}}\xspace}

\begin{titlepage}
\PHyear{2021}       
\PHnumber{079}      
\PHdate{10 May}  

\title{Measurement of the production cross section of prompt $\Xi^0_{\rm c}$ baryons at midrapidity in pp collisions at $\sqrt{s}$ = 5.02 TeV}
\ShortTitle{$\Xi^0_{\rm c}$ production in pp collisions at $\sqrt{s}$ = 5.02 TeV}   

\Collaboration{ALICE Collaboration\thanks{See Appendix~\ref{app:collab} for the list of collaboration members}}
\ShortAuthor{ALICE Collaboration} 

\begin{abstract}
The transverse momentum ($\pt$) differential cross section of the charm-strange baryon $\Xi^0_{\rm c}$
is measured at midrapidity ($|y|<$ 0.5) via its
semileptonic decay into ${\rm e^{+}}\Xi^{-}\nu_{\rm e}$ in pp collisions at $\sqrt{s}$~=~5.02~TeV with the ALICE detector at the LHC.
The ratio of the \pt-differential $\Xi^0_{\rm c}$-baryon and ${\rm D^0}$-meson production cross sections is also reported.
The measurements are compared with simulations with different tunes of the PYTHIA~8 event generator, with predictions from a statistical hadronisation model (SHM) with a largely augmented set of charm-baryon states beyond the current lists of the Particle Data Group, and with models including hadronisation via quark coalescence.
The \pt-integrated cross section of prompt $\Xi^0_{\rm c}$-baryon production at midrapidity is also reported, which is used to calculate the baryon-to-meson ratio $\Xi^0_{\rm c}/{\rm D^0} = 0.20 \pm 0.04~{\rm (stat.)} ^{+0.08}_{-0.07}~{\rm (syst.)}$.
These results provide an additional indication  of a modification of the charm fragmentation from {$\rm e^+e^-$} and {$\rm e^{-}p$} collisions to pp collisions.

\end{abstract}
\end{titlepage}

\setcounter{page}{2} 


\section{Introduction} \label{sec1} 

Measurements of the production of heavy-flavour hadrons (i.e.~containing charm or beauty quarks) in high-energy hadronic collisions provide important tests of quantum chromodynamics (QCD) because perturbative techniques are applicable down to low transverse momentum (\pt) thanks to the large masses of charm and beauty quarks compared to the QCD scale parameter ($\Lambda_{\rm QCD}~\sim$ 200 MeV).
The production cross sections of
heavy-flavour hadrons can be calculated using the factorisation approach~\cite{Collins:1985gm} as a convolution of three factors:
the parton distribution functions (PDFs) of the incoming protons, the hard-scattering cross section at partonic
level, which can be calculated perturbatively in powers of the strong coupling constant $\alpha_{\rm s}$, and the fragmentation function, which parametrises the non-perturbative transition of a heavy quark into a given species of heavy-flavour hadron.
The measurements of D- and B-meson production cross sections at midrapidity in proton--proton (pp) collisions at several centre-of-mass energies at the LHC~\cite{Acharya:2019mgn,Acharya:2017jgo,Abelev:2012vra,Sirunyan:2017xss,Khachatryan:2011mk,Chatrchyan:2011pw,Chatrchyan:2011vh,Acharya:2021cqv} are described within uncertainties by perturbative calculations at next-to-leading order with next-to-leading-log resummation, such as the general-mass variable-flavour-number scheme (GM-VFNS~\cite{Kramer:2017gct,Helenius:2018uul,Kniehl:2020szu}) and the fixed-order next-to-leading-log (FONLL~\cite{Cacciari:1998it, Cacciari:2012ny}) frameworks, over a wide range of \pt. Both calculations use fragmentation functions based on measurements in positron--electron ($\rm{e^+e^-}$) collisions. 

Measurements of the production cross sections of different charm-hadron species, comparing in particular baryon and meson production in various collision systems and centre-of-mass energies, provide insight into the properties of the fragmentation process. Measurements of \mbox{$\Lambda^{+}_{\rm c}$-baryon} production at midrapidity in pp collisions at $\sqrt{s}$ = 5.02, 7, and 13 TeV were reported by the ALICE and CMS collaborations in Refs.~\cite{Sirunyan:2019fnc,Acharya:2017kfy,Acharya:2020uqi,Acharya:2020lrg, SigmacLambdac}. A clear decreasing trend of the $\Lambda^{+}_{\rm c}/{\rm D^0}$ ratio with increasing \pt is seen. The $\Lambda^{+}_{\rm c}/{\rm D^0}$ ratio is measured to be substantially larger than previous measurements at lower centre-of-mass energies in ${\rm e^{+}e^{-}}$ \cite{Albrecht:1988an,Avery:1990bc,Gladilin:2014tba} and electron--proton ({$\rm e^{-}p$}) collisions~\cite{Chekanov:2005mm,Abramowicz:2013eja}, suggesting that the charm fragmentation is not universal among different collision systems. Similar indications were obtained from the measurements of $\Xi^{0,+}_{\rm c}$-baryon and $\Sigma^{0,++}_{\rm c}$-baryon production at midrapidity in pp collisions at $\sqrt{s}$ = 7 and 13 TeV~\cite{Acharya:2017lwf,xic13tev, SigmacLambdac}.

The charm-baryon production cross sections measured at the LHC are substantially larger than the predictions of GM-VFNS calculations and of the POWHEG next-to-leading-order (NLO) generator matched to PYTHIA~6 for the parton shower and the hadronisation stages~\cite{Frixione:2007nw,Kniehl:2020szu,Acharya:2020uqi}. Predictions from QCD-inspired event generators like PYTHIA~8 with Monash tune~\cite{Skands:2014pea}, DIPSY~\cite{Bierlich:2015rha}, and HERWIG 7~\cite{Bahr:2008pv} also underestimate the baryon-to-meson ratios measured at midrapidity.
On the other hand, PYTHIA~8 simulations with tunes including string formation beyond the leading-colour approximation~\cite{Christiansen:2015yqa} qualitatively describe the measured $\Lambda^{+}_{\rm c}/{\rm D^0}$ and $\Sigma^{0,+,++}_{\rm c}/{\rm D^0}$ ratios~\cite{Acharya:2020uqi,SigmacLambdac}, but underestimate the $\Xi^{0,+}_{\rm c}/{\rm D^0}$ ratio~\cite{Acharya:2017lwf,xic13tev}. 
Calculations with a statistical hadronisation model (SHM)~\cite{He:2019tik} based on charm-hadron states listed by the Particle Data Group (PDG)~\cite{Zyla:2020zbs} underestimate the measured baryon-to-meson ratios. The $\Lambda^{+}_{\rm c}/{\rm D^0}$ and $\Sigma^{0,+,++}_{\rm c}/{\rm D^0}$ ratios are qualitatively described by the SHM calculations if a larger set of yet-unobserved higher-mass charm-baryon states is considered under the guidance of the relativistic quark model (RQM)~\cite{Ebert:2011kk} and of lattice QCD~\cite{Briceno:2012wt}. However, the $\Xi^{0,+}_{\rm c}/{\rm D^0}$ ratios are still underestimated with the inclusion of the additional baryonic states~\cite{xic13tev}. An enhancement of the charmed baryon-to-meson ratio is expected also by models employing  hadronisation of charm quarks via recombination in pp collisions~\cite{Song:2018tpv,Minissale:2020bif}.
In the quark (re-)combination mechanism (QCM) model~\cite{Song:2018tpv}, the charm quark is combined with a co-moving light antiquark or with two co-moving quarks to form a charmed meson or baryon. The Catania model~\cite{Minissale:2020bif} implements charm-quark hadronisation via both coalescence, implemented via Wigner formalism~\cite{Dover:1991zn}, and fragmentation. 

Finally, it should be noted that an increased yield of charmed baryons ($\Lambda_{\rm c}$, $\Xi_{\rm c}$, $\Omega_{\rm c}$) has significant consequences on the determination of the total charm cross section in pp collisions at the LHC~\cite{Acharya:2017jgo,Abelev:2012vra}. 
In the context of the heavy-ion programme at the LHC, the $\rm c\overline{c}$ production cross section per nucleon--nucleon collision is a fundamental ingredient for the determination of the amount of charmonium production by (re)generation in a quark--gluon plasma (QGP), a mechanism that is supported by J/$\psi$ measurements in nucleus--nucleus collisions at the LHC~\cite{Adam:2016rdg,Acharya:2019lkh}. A precise determination of the $\rm c\overline{c}$ production cross section in pp collisions at midrapidity will offer a stronger constraint to models implementing J/$\psi$ regeneration in the QGP~\cite{Andronic:2019wva,Zhao:2011cv,Liu:2009nb}.
In addition, measurements of open heavy-flavour baryon production in heavy-ion collisions provide a unique information on hadronisation mechanisms in the QGP. 
Models implementing charm-quark hadronisation via coalescence in addition to fragmentation~\cite{Lee:2007wr,Oh:2009zj,Plumari:2017ntm} predict an enhanced baryon-to-meson ratio in heavy-ion collisions with respect to pp collisions.

In this article, we report the measurement of the \pt-differential production cross section of prompt $\Xi^0_{\rm c}$ baryons in pp collisions at $\sqrt{s}$ = 5.02 TeV and its ratio to the measured production cross section of prompt ${\rm D^0}$ mesons (i.e.\ produced directly in the hadronisation of charm quarks and in the decays of directly produced excited charm states)~\cite{Acharya:2019mgn,Acharya:2021cqv}. The $\Xi^0_{\rm c}$ baryons and their antiparticles are reconstructed at midrapidity ($|y|~<$~0.5) in the transverse momentum interval 2 $< \pt <$ 8~GeV/$c$ via the semileptonic decay mode $\Xi^0_{\rm c} \rightarrow {\rm e}^{+}\Xi^{-}\nu_{\rm e}$ and its charge conjugate. We have recently constrained the absolute branching ratio (BR) of this $\Xi^0_{\rm c}$ decay in pp collision at $\sqrt{s}$ = 13 TeV by measuring the $\Xi^0_{\rm c}$ production via two different decay channels, $\Xi^0_{\rm c} \rightarrow {\rm e}^{+}\Xi^{-}\nu_{\rm e}$ and $\Xi^0_{\rm c} \rightarrow {\pi}^{+}\Xi^{-}$~\cite{xic13tev}.
This BR value is used in this analysis and it is also used to update the previously published measurement of inclusive $\Xi^0_{\rm c}$ \pt-differential cross section in pp collisions at $\sqrt{s}$ = 7 TeV~\cite{Acharya:2017lwf}.

The article is organised as follows. Section \ref{sec2} describes the experimental setup, focusing on the detectors employed in the analysis and the data-taking conditions. The analysis details and the estimation of the systematic uncertainties are discussed in Sec.~\ref{sec3}. Section \ref{sec4} presents the results, namely the \pt-differential production cross section of prompt $\Xi^0_{\rm c}$ baryons and the $\Xi^0_{\rm c}/{\rm D^0}$ cross-section ratio, which are compared with different model calculations.  The \pt-integrated production cross section of prompt $\Xi^0_{\rm c}$ baryons, and the corresponding
$\Xi^0_{\rm c}/{\rm D^0}$ ratio, extrapolated down to \pt = 0,  are also reported and compared with model calculations.
Finally, conclusions are drawn in Sec.~\ref{sec6}.

\section{Experimental apparatus and data sample}\label{sec2}

The ALICE apparatus, described in detail in Refs.~\cite{Aamodt:2008zz,bib::Alice_performance}, consists of a central barrel covering the pseudorapidity region $|\eta|<0.9$ placed inside a solenoidal magnet that provides a $B$ = 0.5 T field parallel to the beam direction, a muon spectrometer at forward pseudorapidity ($-4<\eta< -$2.5), and a set of detectors at forward/backward rapidity for triggering and event selection.
The detectors used for reconstruction and identification of the $\Xi^0_{\rm c}$ decay products are the Inner Tracking System (ITS)~\cite{1748-0221-5-03-P03003}, the Time Projection Chamber (TPC)~\cite{Alme:2010ke},  and the Time-Of-Flight detector (TOF)~\cite{tofperf}. 

The ITS consists of six cylindrical layers of silicon detectors.
The two innermost layers, equipped with Silicon Pixel Detectors (SPD), provide a space-point position resolution of 12 $\mu$m and 100 $\mu$m in the $r\varphi$ and the beam direction, respectively. The third and fourth layers consist of Silicon Drift Detectors (SDD), while the two outermost layers are equipped with Silicon Strip Detectors (SSD).

The TPC is the main tracking detector in the central barrel. With up to 159 space points to reconstruct the charged-particle trajectory,  it provides  charged-particle momentum measurement together with excellent two-track separation and particle identification via d$E$/d$x$ determination with a resolution better than 5\%~\cite{Alme:2010ke}.

The TOF detector provides the measurement of the flight time of charged particles from the interaction point to the detector radius of 3.8 m, with an overall resolution of about 80 ps. The collision time is obtained using either the information from the T0 detector~\cite{Bondila:2005xy}, or the TOF detector, or a combination of the two. The T0 detector consists of two arrays of {\v C}erenkov counters, located on both sides of the interaction point, covering the pseudorapidity intervals $-3.28<\eta<- 2.97$ and $4.61<\eta<4.92$. 

The V0 detector~\cite{Abbas:2013taa}, composed of two arrays of 32 scintillators each, covering the pseudorapidity intervals $-3.7<\eta<-1.7$ and $2.8<\eta<5.1$, provides the minimum bias (MB) trigger used to collect the data sample. In addition, the timing information of the two V0 arrays and the correlation between the number of hits and track segments in the SPD were used for an offline event selection, in order to remove background due to interactions between one of the beams and the residual gas present in the beam vacuum tube. 

In order to maintain a uniform acceptance in pseudorapidity, collision vertices were required to be within
$\pm$10 cm from the centre of the detector along the beam line direction. The pile-up events (less than 1\%) were rejected by detecting multiple primary vertices using track segments defined with the SPD layers. After the aforementioned selections, the data sample used for the analysis consists of about 990 million MB events, corresponding to an integrated luminosity of $\mathcal{L}_{\rm int}$~=~(19.3~$\pm$~0.4)~${\rm nb^{-1}}$~\cite{ALICE-PUBLIC-2018-014}, collected during the 2017 pp run at $\sqrt{s}$ = 5.02 TeV.

\section{Data analysis}  \label{sec3}

The analysis is performed using similar techniques to those reported in Ref.~\cite{Acharya:2017lwf}.
The $\Xi^0_{\rm c}$ baryons are reconstructed via the semileptonic decay mode  $\Xi^0_{\rm c} \rightarrow {\rm e}^{+}\Xi^{-}\nu_{\rm e}$, and its charge conjugate. The $\Xi^0_{\rm c}$ candidates are defined from ${\rm e}^{+}\Xi^{-}$ pairs formed by combining positrons and $\Xi^{-}$ baryons. The $\Xi^{0}_{\rm c}$ raw yield is obtained by counting the ${\rm e}^{+}\Xi^{-}$ pairs in $p_{\rm T}^{\rm e\Xi}$ intervals, where $p_{\rm T}^{\rm e\Xi}$ is the transverse momentum of the ${\rm e}^{+}\Xi^{-}$ pair, after subtracting the combinatorial background, as described in Sec.~\ref{eXiBkg}.  The $p_{\rm T}^{\rm e\Xi}$ distribution of ${\rm e}^{+}\Xi^{-}$ pairs is corrected for the missing momentum of the neutrino using unfolding techniques, in order to obtain the $\Xi^{0}_{\rm c}$ raw yield in intervals of $\Xi^{0}_{\rm c}$ \pt, as described in Sec.~\ref{unfolding}. The contribution of $\Xi^0_{\rm c}$ baryons originating from beauty-hadron decays is subtracted from the measured yield by using perturbative quantum chromodynamics (pQCD) calculations of the beauty-quark cross section together with the fragmentation fractions of beauty quarks into hadrons measured by LHCb~\cite{Aaij:2019pqz}, and the acceptance and efficiency values estimated from simulations as described in Sec.~\ref{feeddown}.
Charge conjugate modes are implied everywhere, unless otherwise stated. The final results are obtained as the average of particles and antiparticles.

\subsection{Reconstruction of $\rm e^{\pm}$ and $\Xi^{\pm}$ candidates}\label{RecoeXi} 
\begin{figure}[tb!]
\begin{minipage}[c]{0.5\linewidth}
\includegraphics[width=.9\textwidth]{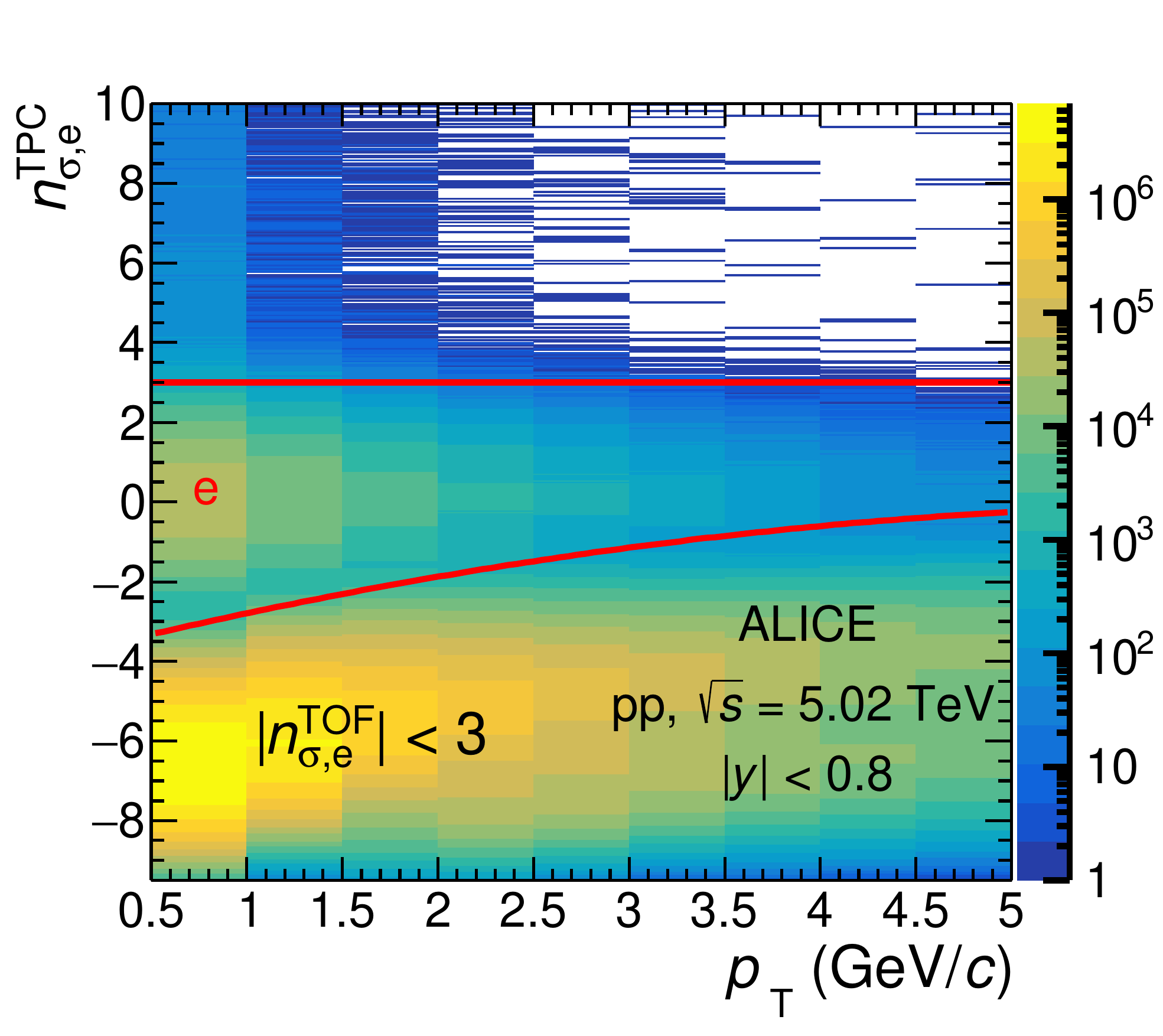}
\end{minipage}
\hspace{0.5cm}
\begin{minipage}[c]{0.5\linewidth}
\includegraphics[width=.9\textwidth]{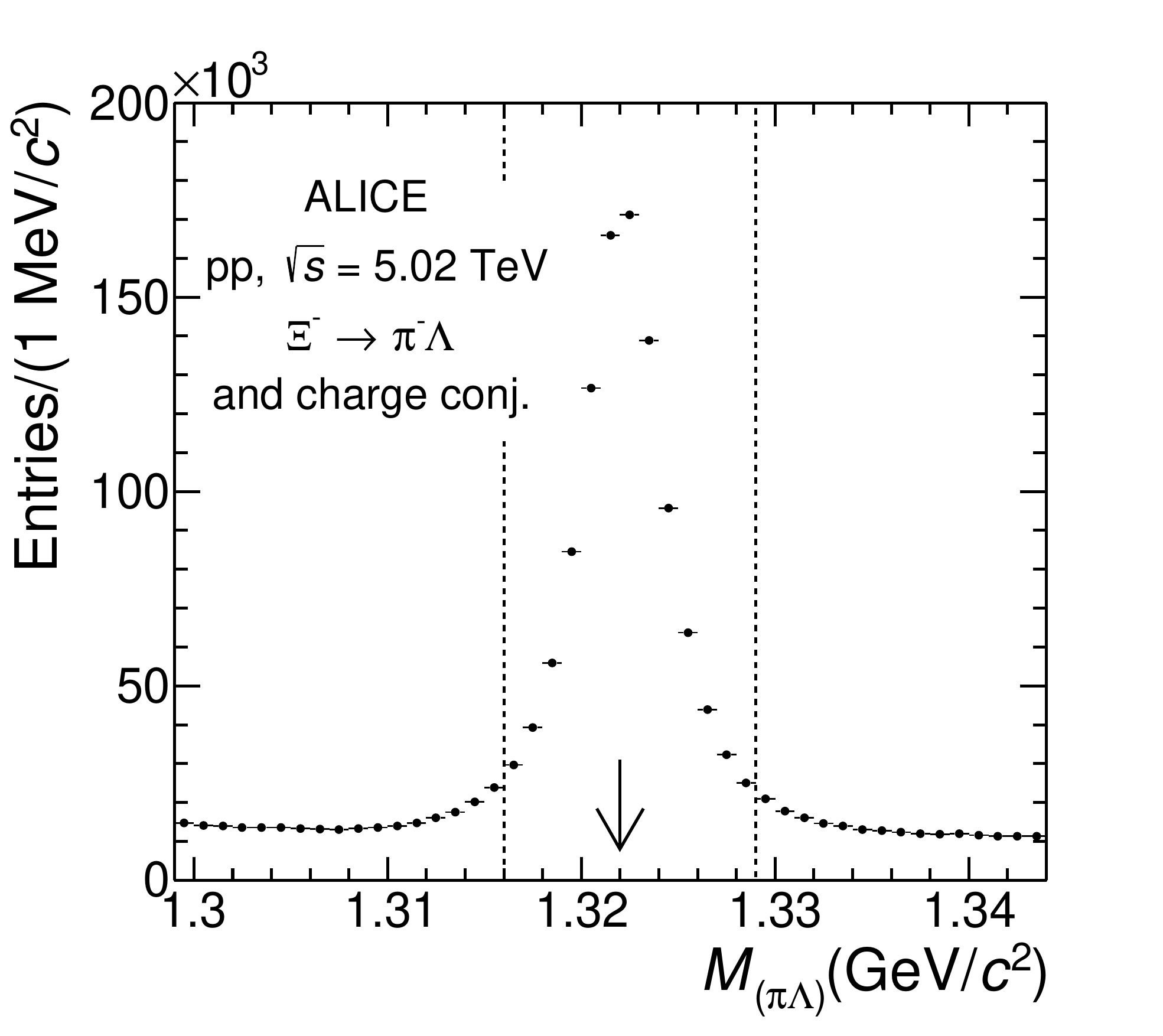}
\end{minipage}%
\caption{Left panel: $n_{\sigma\textrm{,e}}^\textrm{TPC}$ distribution as a function of the electron \pt after applying the particle identification criteria on the TOF signal (see text for details). Right panel: invariant mass distribution of $\Xi^{-} \rightarrow \pi^{-}\Lambda$ (and charge conjugate) candidates integrated over $ p_{\rm T}^{\Xi^{-}}$.
The arrow indicates the world average $\Xi^{-}$ mass~\cite{Zyla:2020zbs} and the dashed lines define the interval in which the $\Xi^{-}$ candidates are selected for the $\Xi^0_{\rm c}$  reconstruction (see text for details).}
\label{fig1}
\end{figure}

Candidate electron and positron tracks satisfying $|\eta|<$~0.8 and $\pt > 0.5$~GeV/$c$ are required to have a number of crossed TPC pad rows larger than 80, a $\chi^2$ normalised to the number of associated TPC clusters smaller than 4,  and at least 3 hits in the ITS. These selection criteria suppress the contribution from short tracks, which are unlikely to originate from the primary vertex. In order to reject electrons from photon conversions
occurring in the detector material outside the innermost SPD layer,  the electron candidate tracks are required to have
associated hits in the two SPD layers of the ITS~\cite{Acharya:2018upq,Acharya:2019mom}. In addition, at least 50 TPC clusters are required for the calculation of the d$E$/d$x$ signal. Electrons are identified using the
d$E$/d$x$ and the time-of-flight measurements in the TPC and TOF detectors. The selection is applied on the $n_{\sigma\textrm{,e}}^\textrm{TPC}$ and $n_{\sigma\textrm{,e}}^\textrm{TOF}$ variables defined as the difference between the measured
d$E$/d$x$ or time-of-flight values and the ones expected for electrons, divided by the corresponding detector resolution. 
In the left panel of Fig.~\ref{fig1}, the $n_{\sigma\textrm{,e}}^\textrm{TPC}$ distribution as a function of the candidate electron \pt is shown for tracks with a time-of-flight compatible with the value expected for an electron within $|n_{\sigma\textrm{,e}}^\textrm{TOF}|~<$~3.
The following criterion is applied on the TPC d$E$/d$x$ signal to select electron candidates: $-$~3.9~+~1.2\pt~$-$~0.094$\pt^2~<~|n_{\sigma\textrm{,e}}^\textrm{TPC}(\pt)|~<$~3 (with \pt in units of GeV/$c$), which is represented by the red lines in the left panel of Fig.~\ref{fig1}. The \pt-dependent lower limit on $|n_{\sigma\textrm{,e}}^\textrm{TPC}|$ is optimised to reject hadrons. An electron purity of 98\% is achieved over the whole \pt range.

Further rejection of background electrons originating from Dalitz decays of neutral mesons and photon
conversions in the detector material (``photonic'' electrons) is obtained using a technique based on the invariant mass of $\rm e^+e^-$ pairs~\cite{Acharya:2019lkh,Acharya:2019lkw}. The electron (positron) candidates are paired with opposite-sign tracks
from the same event passing loose identification criteria ($|n_{\sigma\textrm{,e}}^\textrm{TPC}|~<$ 5 without any TOF requirement) and are rejected if they form at least one $\rm e^{+}e^{-}$ pair with an invariant mass smaller than 50 MeV/$c^2$.
Loose electron identification criteria are used in order to have a high  efficiency of finding the partners~\cite{Adam:2015qda}.  With this selection the fraction of signal lost due to mistagging is less than 2\%, as discussed in Sec.~\ref{unfolding}.

The $\Xi^{-}$ baryons are reconstructed from the decay chain $\Xi^{-} \rightarrow \Lambda\pi^{-}$ ($\rm BR=99.887\pm 0.035\%$), followed by $\rm \Lambda \rightarrow p\pi^{-}$ ($\rm BR=63.9\pm 0.5\%$)~\cite{Zyla:2020zbs}. Tracks used to define $\Xi^{-}$ candidates are required to have a number of crossed TPC pad rows larger than 70 and a d$E$/d$x$ signal in the TPC consistent with the expected value for protons (pions) within 4$\sigma$. The $\Xi^{-}$ and $\Lambda$ baryons have long lifetimes ($c\tau$ of about 4.91 cm and 7.89 cm, respectively~\cite{Zyla:2020zbs}), and thus they can be selected exploiting their characteristic decay topologies~\cite{Acharya:2019kyh}. Pions originating directly from $\Xi^{-}$ decays are selected by requiring a minimum distance of closest approach ($d_{\rm 0}$) of their tracks to the primary vertex, $d_{\rm 0}~>$ 0.05 cm, while protons and pions originating from $\Lambda$ decays are required to have $d_{\rm 0}~>$ 0.07 cm. 
The $d_{\rm 0}$ of the $\Lambda$ trajectory to the primary vertex is required to be larger than 0.05 cm, while the cosine of the $\Lambda$ pointing angle, which is the angle between the reconstructed $\Lambda$ momentum and the line connecting the $\Lambda$ and $\Xi^{-}$ decay vertices, is required to be larger than 0.98. The cosine of the pointing angle of the reconstructed $\Xi^{-}$ momentum to the primary vertex is required to be larger than 0.983. The radial distances of the $\Xi^{-}$ and $\Lambda$ decay vertices from the beam line are required to be larger than 0.4 and 2.7 cm, respectively.
These selection criteria are tuned to reduce the background and enhance the purity of the signal. In the right panel of Fig.~\ref{fig1} the $\Xi^{-}$ peak in the $\pi^{-}\Lambda$ invariant mass distribution integrated for $\pt^{\Xi^{-}} > 0$ is shown. Only
$\Xi^{-}$ candidates with invariant masses within 8 MeV/$c^{2}$
from the world average $\Xi^{-}$ mass (1321.71~$\pm$~0.07 MeV/$c^{2}$~\cite{Zyla:2020zbs}),
indicated by an arrow in the right panel of Fig.~\ref{fig1}, are kept for further analysis.

\subsection{ Analysis of ${\rm e^{\pm}}\Xi^{\mp}$ invariant mass distribution}\label{eXiBkg}

\begin{figure}[ht!]
\begin{minipage}[c]{0.5\linewidth}
\includegraphics[width=.9\textwidth]{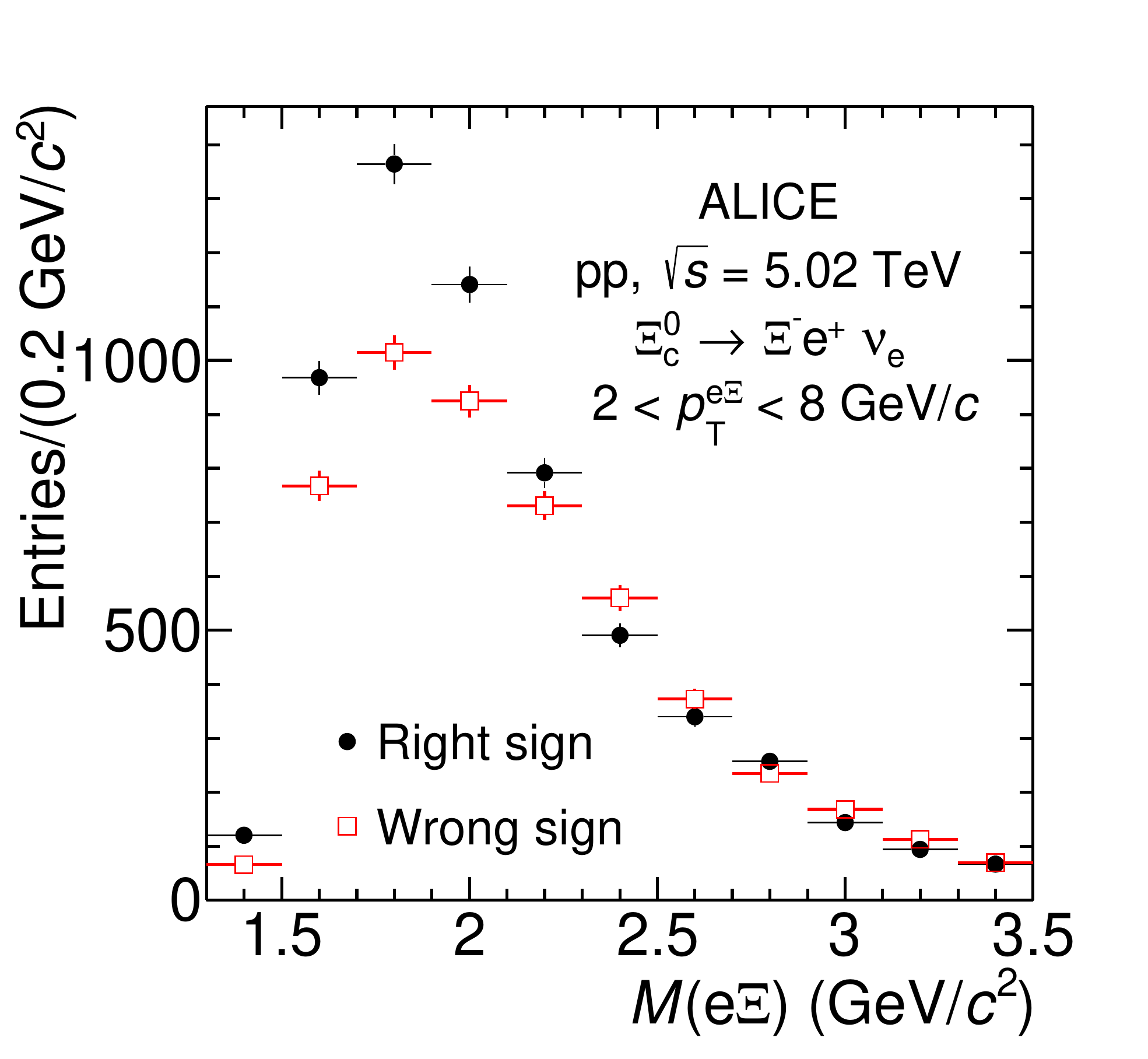}
\end{minipage}
\hspace{0.3cm}
\begin{minipage}[c]{0.2\linewidth}
\includegraphics[width=2.2\textwidth]{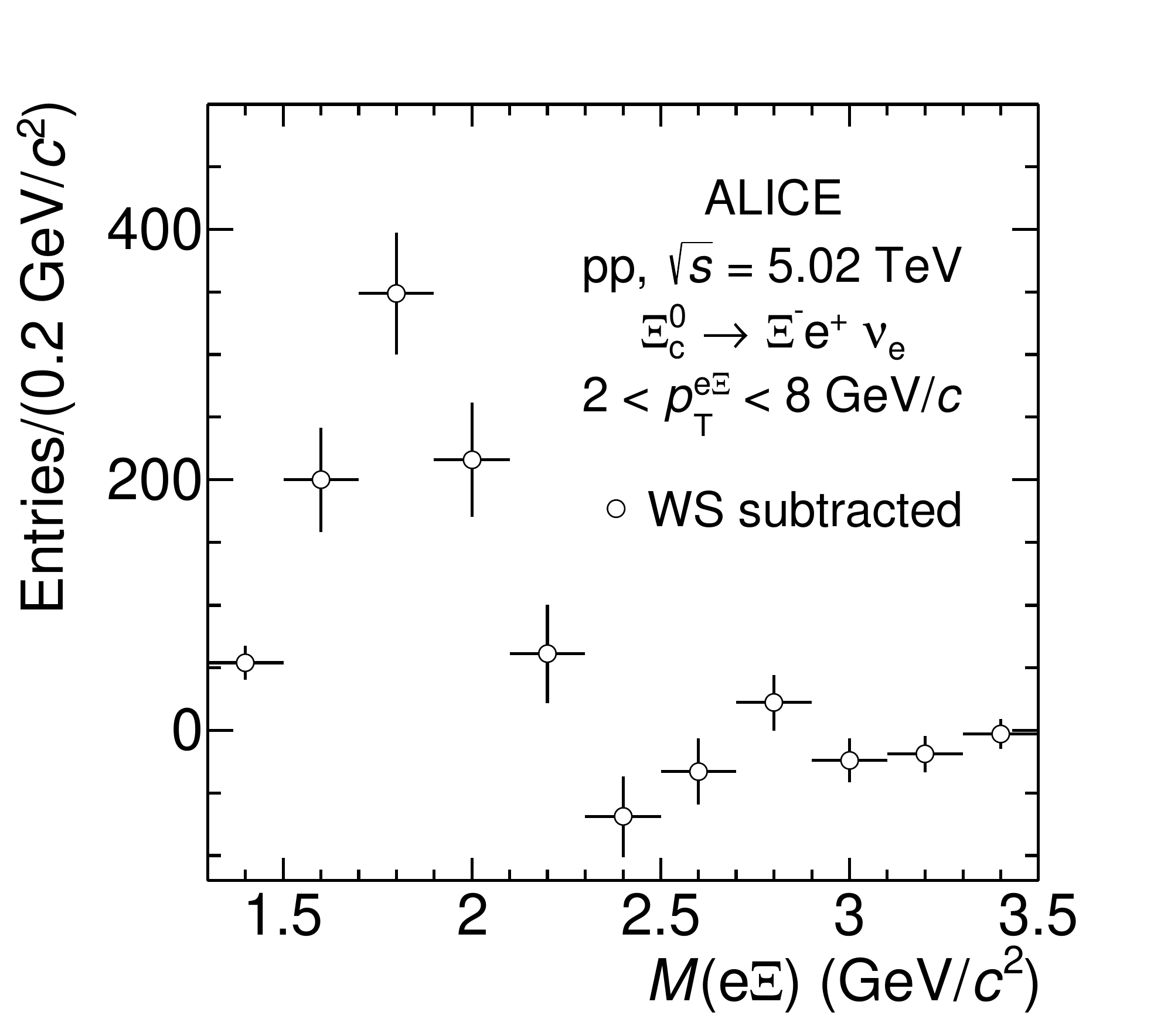}
\end{minipage}%
\caption{Left panel: invariant mass distributions of right-sign and wrong-sign ${\rm e}\Xi$ pairs with 2 $< \pt <$ 8 GeV/$c$. Right panel: invariant mass distribution of $\Xi^0_{\rm c}$ candidates obtained by subtracting the wrong-sign pair yield from the right-sign pair yield.
}
\label{fig2}
\end{figure}

The $\Xi^0_{\rm c}$ candidates are defined from  ${\rm e}^{+}\Xi^{-}$ pairs. Only pairs with an opening angle smaller than 90 degrees are used for the analysis. Due to the undetected neutrino, the invariant mass distribution of ${\rm e}^{+}\Xi^{-}$ pairs does not show a peak at the $\Xi^0_{\rm c}$ mass. Following the same approach adopted and described in Ref.~\cite{Acharya:2017lwf}, the background contributions are estimated exploiting the fact that $\Xi^0_{\rm c}$ baryons and their antiparticles decay only into ${\rm e}\Xi$ pairs with opposite charge sign (${\rm e}^{+}\Xi^{-}$ and ${\rm e}^{-}\overline{\Xi}^{+}$), 
denoted as right-sign (RS), and not into same-sign pairs (${\rm e}^{-}\Xi^{-}$ and ${\rm e}^{+}\overline{\Xi}^{+}$), denoted as wrong-sign (WS), while combinatorial background candidates contribute equally to both RS and WS pairs. The $\Xi^0_{\rm c}$ raw yield is obtained from the invariant mass distribution of RS pairs after subtracting the WS contribution. Other contributions to e$\Xi$ pairs, such as those from $\Xi_{\rm b}^{0,-}$ semileptonic decays
to WS pairs, which do not give rise to RS pairs, are corrected for after the subtraction, as described in Sec.~\ref{unfolding}.
In the left panel of Fig.~\ref{fig2} the uncorrected invariant mass distributions of WS and RS pairs in the interval 2~$<~\pt^{{\rm e}\Xi}~<$~8~GeV/$c$ are shown for illustration. In the right panel of Fig.~\ref{fig2} the invariant mass distribution of $\Xi^0_{\rm c}$ candidates obtained after subtracting the WS pair yield from the RS yield is shown. Only ${\rm e}^{+}\Xi^{-}$ pairs satisfying $m_{{\rm e}\Xi}~<$~2.5~GeV/$c^2$ are considered.

\subsection{Corrections and unfolding}\label{unfolding}
The raw yield obtained by counting the ${\rm e}^{+}\Xi^{-}$ candidates in bins of $p_{\rm T}^{{\rm e}\Xi}$  after the subtraction of the WS pairs needs to be corrected for the signal loss due to mistagging of photonic electrons, and for the $\Xi_{\rm b}^{0,-}$ contribution in the WS pairs. Finally, the $p_{\rm T}^{{\rm e}\Xi}$-differential spectrum is corrected for the missing neutrino momentum to obtain the $\Xi^{0}_{\rm c}$ raw yield in intervals of $\Xi^{0}_{\rm c}$ \pt. 

The probability of wrongly tagging an electron as photonic is estimated by applying the tagging algorithm, described in Sec.~\ref{RecoeXi}, to ${\rm e^{+}e^{+}}$ and ${\rm e^{-}e^{-}}$ pairs. The resulting correction is smaller than 2\%, with a mild dependence on the \pt of the ${\rm e}^{+}\Xi^{-}$ pair, as it was also observed in Refs.~\cite{Acharya:2017lwf,xic13tev}.

Decays of $\Xi_{\rm b}^{0,-}$ to electrons, $\Xi_{\rm b}^{0,-} \rightarrow  {\rm H}_{\rm c} {\rm e^{-}}{\rm X}$ (where ${\rm H}_{\rm c}$ is any charmed baryon), followed by $\rm H_{\rm c}$ decays to $\Xi^{-}$, ${\rm H}_{\rm c} \rightarrow \Xi^{-}{\rm X}$, contribute to the WS invariant mass distribution and not to the RS one, giving rise to a background over-subtraction. In order to estimate this contribution, assumptions must be made for the branching ratio of $\Xi^{0,-}_{\rm b}$ into ${\rm e}^-\Xi^-\bar{\nu}_{\rm e}{\rm X}$ and for the $\Xi_{\rm b}^{0,-}$ production cross sections, which are not measured.
First, the shape of the transverse momentum distribution of $\Xi^{0,-}_{\rm b}$ baryons is assumed to be the same as that of $\Lambda^0_{\rm b}$ baryons. The CMS collaboration reported a measurement of the \pt-differential $\Lambda^0_{\rm b}$ production cross section multiplied by the ${\rm BR}(\Lambda^0_{\rm b} \rightarrow {\rm J}/\psi \Lambda)$ in pp collisions at $\sqrt{s}$ = 7 TeV~\cite{Chatrchyan:2012xg}. To scale the $\Lambda^0_{\rm b}$ measurement at the centre-of-mass energy of 5.02 TeV,  the ratio of the beauty-hadron cross sections at \mbox{$\sqrt{s}$ = 7 TeV} and 5.02 TeV obtained with FONLL pQCD calculations is used~\cite{Cacciari:1998it, Cacciari:2012ny}. The second assumption is that the fraction of beauty quarks that hadronise into $\Lambda^0_{\rm b}$ and $\Xi^{0,-}_{\rm b}$ are the same as those in $\rm e^+e^-$ collisions. The yield of $\Xi^{0,-}_{\rm b} \rightarrow {\rm e}^-\Xi^-\bar{\nu}_{\rm e}{\rm X}$ is therefore computed using  i) the $\sqrt{s}$-scaled $\Lambda^0_{\rm b}$ cross section, ii) the values of $f({\rm b} \rightarrow \Xi^{0,-}_{\rm b})\times {\rm BR}(\Xi^{0,-}_{\rm b} \rightarrow {\rm e}^-\Xi^-\bar{\nu}_{\rm e}{\rm X} $)~\cite{Zyla:2020zbs}  and $f({\rm b} \rightarrow \Lambda^0_{\rm b})\times {\rm BR}(\Lambda^0_{\rm b} \rightarrow {\rm J}/\psi \Lambda)$~\cite{Barate:1997if} measured in  ${\rm e^{+}e^{-}}$ collisions, and iii) the $\Xi^{0,-}_{\rm b} \rightarrow {\rm e}^-\Xi^-\bar{\nu}_{\rm e}{\rm X}$ acceptance $\times$ efficiency $({\rm Acc} \times\varepsilon)$ from the simulations described below. The contribution to the WS pair yield from $\Xi_{\rm b}^{0,-}$ baryon decays is estimated to be about 2\%.

\begin{figure}[tb]
\centering
\includegraphics[width=.48\textwidth]{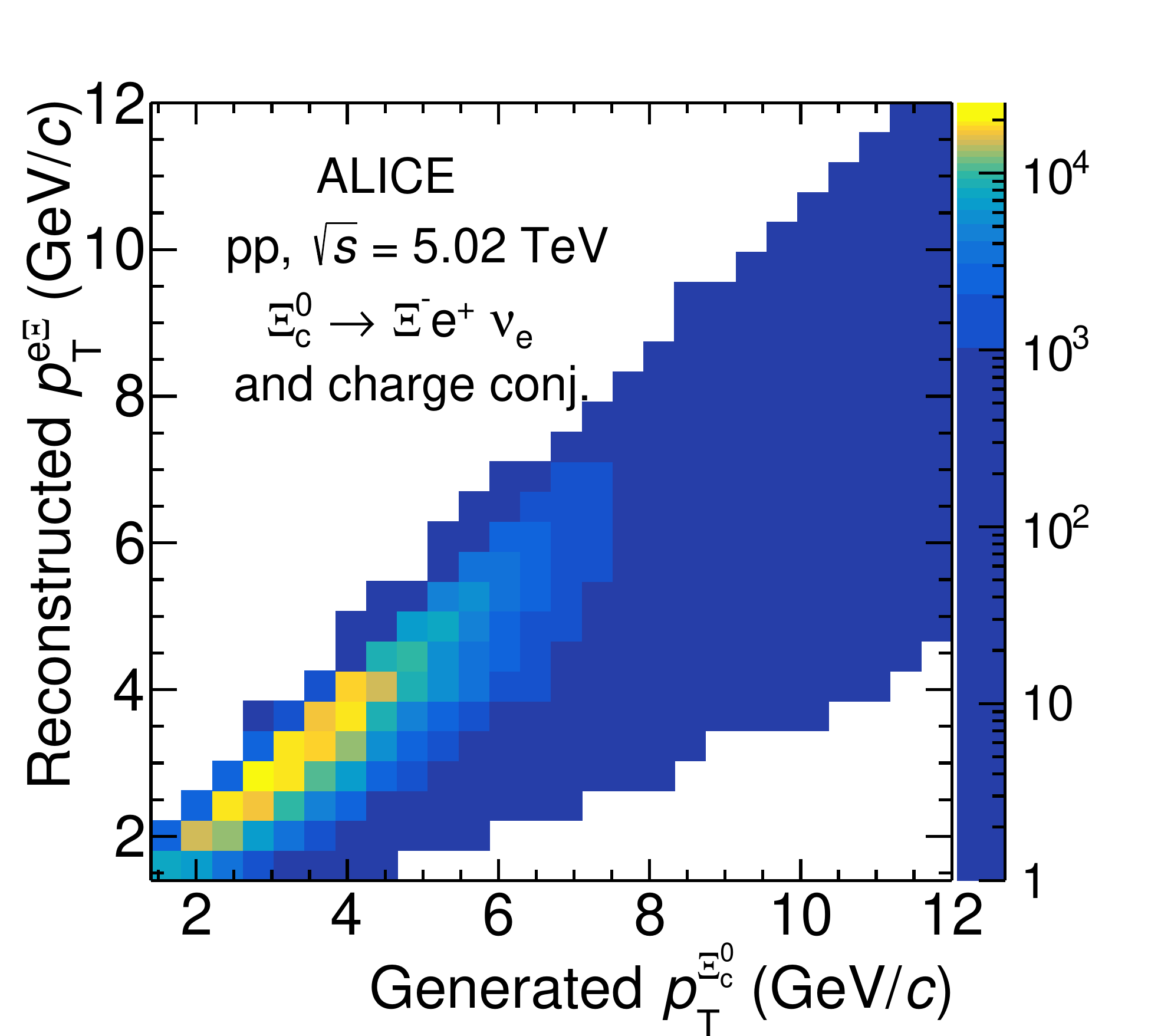}
\caption{Correlation matrix between the generated $\Xi^0_{\rm c}$-baryon \pt and the reconstructed $ {\rm e}^{+}\Xi^{-}$ pair \pt, obtained from the simulation based on PYTHIA 8 described in the text.}
\label{fig3}
\end{figure}

The correction for the missing momentum of the neutrino is performed by using an unfolding technique with a response matrix which represents the correlation between the \pt of the $\Xi_{\rm c}^0$ baryon and that of the reconstructed ${\rm e}^{+}\Xi^{-}$ pair. The response matrix is determined through a simulation with the PYTHIA~8.243 event generator~\cite{PYTHIA8} and the GEANT 3 transport code~\cite{Brun:1082634}, including a realistic description of the detector
conditions and alignment during the data taking period. 
The response matrix needs to be determined using a realistic $\Xi^0_{\rm c}$-baryon \pt distribution which is not known a priori. Therefore, a two-step iterative procedure is adopted. In the first step, the response matrix is obtained with the \pt distribution generated with PYTHIA~8. This matrix is used to calculate a first estimate of the $\Xi^0_{\rm c}$ \pt-differential spectrum from the measured \pt distribution of ${\rm e}^{+}\Xi^{-}$ pairs. The $\Xi^0_{\rm c}$ \pt distribution from this first iteration is used to reweight the response matrix, which is then used for the second iteration. The response matrix obtained from this procedure is shown in Fig.~\ref{fig3}.
The Bayesian unfolding technique~\cite{DAgostini:1994fjx} implemented in the RooUnfold package~\cite{Adye:2011gm} is used. In this analysis the Bayesian procedure required three iterations to converge.
The response matrix used in the unfolding procedure is defined in the transverse
momentum interval $1.4<$\pt$<12$ GeV/$c$, which is wider than the \pt interval used for the cross
section measurement, to avoid edge effects at the lowest and highest \pt intervals of the measurement.

\subsection{Reconstruction efficiency and feed-down subtraction} ~\label{feeddown}
The \pt-differential cross section of prompt $\Xi^0_{\rm c}$-baryon production is obtained as

\begin{equation}
	\frac{\text{d}^2\sigma^{\Xi^0_{\rm c}}}{\text{d}p_\mathrm{T}\text{d}y}=\frac{1}{\rm BR}\times\frac{1}{2\Delta y\Delta p_\mathrm{T}}\times\frac{f_{\rm prompt} \times N^{\Xi^{0}_{\rm c}+\overline{\Xi}^{0}_{\rm c}}_\text{raw}}{({\rm Acc} \times\varepsilon)_{\rm prompt}}\times\frac{1}{\mathcal{L}_{\rm int}},
	\label{eq::corrected_yield}
\end{equation}

where $N^{\Xi^{0}_{\rm c}+\overline{\Xi}^{0}_{\rm c}}_\text{raw}$ is the raw yield after the unfolding correction in a given \pt interval with width $\Delta p_\mathrm{T}$,
$f_{\rm prompt}$ is the fraction of prompt $\Xi^0_{\rm c}$ in the raw yield of $\Xi^0_{\rm c}$, BR is the branching ratio for the considered decay mode, and $\mathcal{L}_{\rm int}$ is the integrated luminosity.  The $({\rm Acc} \times\varepsilon)_{\rm prompt}$ factor is the product of detector acceptance and efficiency for prompt  $\Xi^0_{\rm c}$ baryons, where $\varepsilon$ accounts for the reconstruction and selection of the $\Xi^0_{\rm c}$ decay-product tracks and the $\Xi^0_{\rm c}$-candidate selection. The factor $\Delta y$ represents the width of the rapidity interval in which the generated ${\Xi}^{0}_{\rm c}$ are considered and it is applied to obtain the cross section in one unit of rapidity. The factor 1/2 takes into account that $N^{\Xi^0_{\rm c}}_\text{raw}$ includes both particles and antiparticles, while the cross section is given for particles only. The BR of the considered semileptonic decay channel is calculated from the ratio $\rm BR(\Xi_c^0\rightarrow \Xi^-e^+\nu_e)/\rm BR(\Xi_c^0\rightarrow \Xi^{-}\pi^+)$ = 1.36 $\pm$ 0.14 (stat.) $\pm$ 0.19 (syst.), measured by ALICE in pp collisions at $\sqrt{s}$ = 13 TeV~\cite{xic13tev}, which is multiplied by the hadronic decay branching ratio $\rm BR(\Xi_c^0\rightarrow \Xi^{-}\pi^+)$ reported in the PDG~\cite{Zyla:2020zbs} to get $\rm BR(\Xi_c^0\rightarrow \Xi^-e^+\nu_e)$ = (1.94$\pm$0.55)\%.

\begin{figure}[th!]
\centering
\includegraphics[width=.49\textwidth]{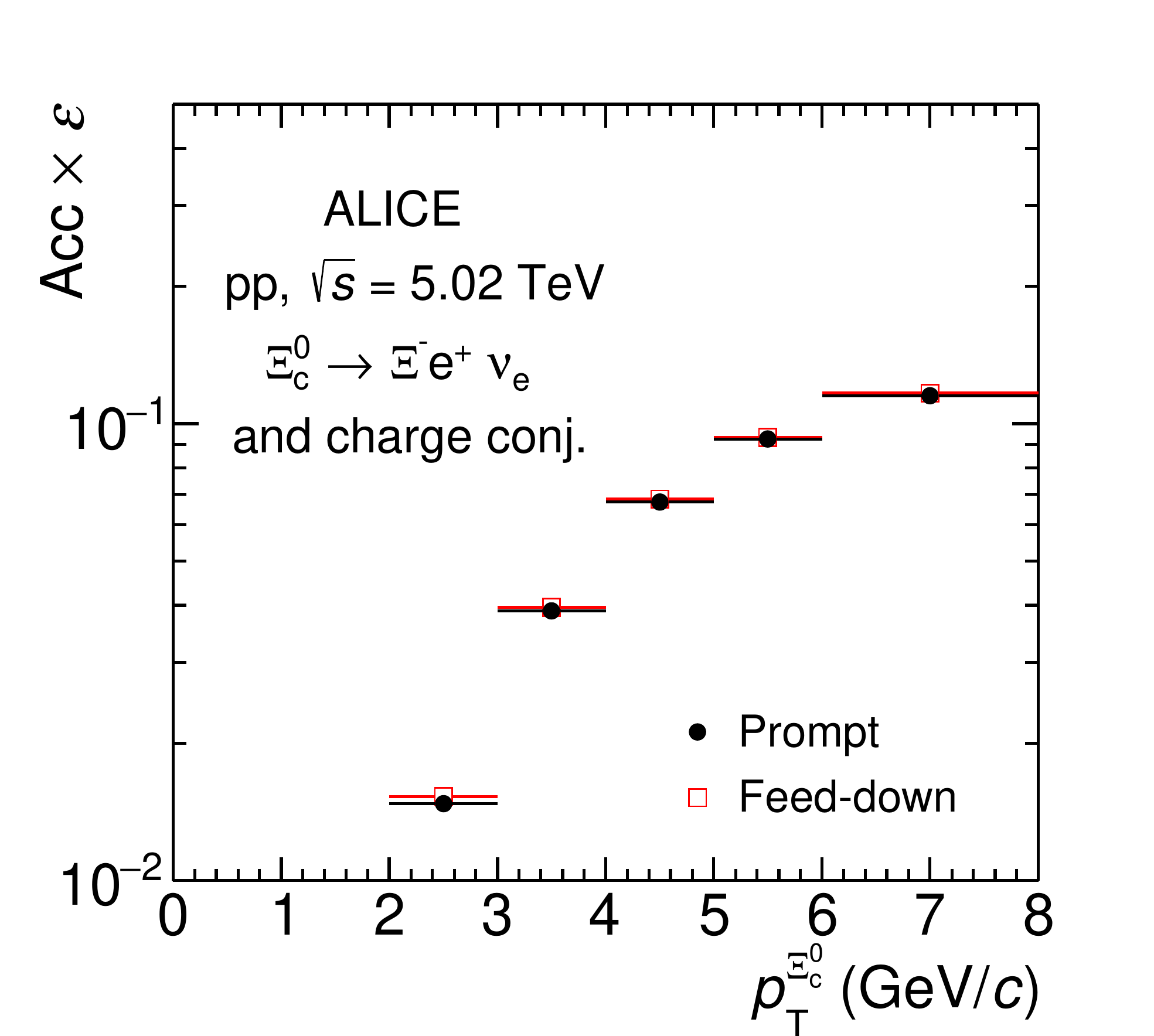}
\includegraphics[width=.50\textwidth]{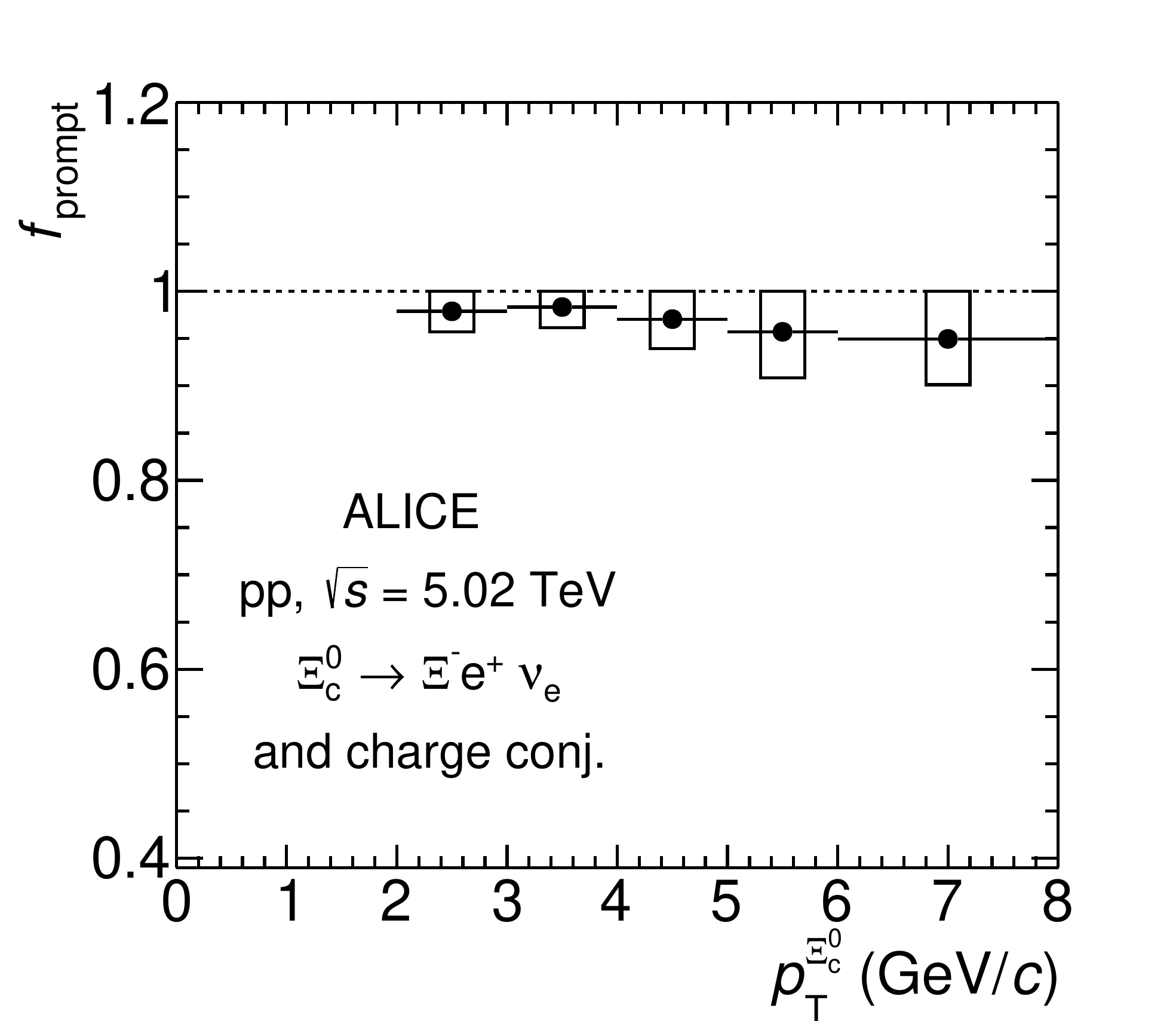}
\caption{Left panel: product of acceptance and efficiency for prompt and feed-down $\Xi^0_{\rm c}$ baryons in pp collisions at $\sqrt{s} =$ 5.02 TeV as a function of
\pt. Right panel: fraction of prompt $\Xi^0_{\rm c}$ baryons in the raw yield ($f_{\rm prompt}$) as a function of \pt. The systematic uncertainties of $f_{\rm prompt}$ are shown as boxes.}
\label{fig4}
\end{figure}

The $({\rm Acc} \times\varepsilon)$ factor is obtained from the same simulations used to determine the response matrix in which the detector and data taking conditions are reproduced. 
The $({\rm Acc} \times\varepsilon)$ is computed separately for prompt and feed-down (produced in beauty-hadron decays) $\Xi^0_{\rm c}$ baryons and is reported in the left panel of Fig.~\ref{fig4}. 
The efficiencies of prompt and feed-down baryons are consistent with each other within uncertainties because the applied selection criteria are not sensitive to the displacement by a few hundred micrometers of the prompt and feed-down $\Xi^0_{\rm c}$ decay vertices from the collision point. 
In order to compute the efficiency with a realistic momentum distribution of $\Xi^0_{\rm c}$ baryons, the \pt shape of the $\Xi^0_{\rm c}$ baryons from the PYTHIA~8 simulation is reweighted to match the measured one via a two-step iterative procedure similar to the one used for the response matrix. 

The factor $f_{\rm prompt}$ is calculated as

\begin{equation}
	f_{\rm prompt}
	= 1 - \frac{N^{ \Xi^0_{\rm c} \ \textnormal {feed-down}}}{N^{\Xi^{0}_{\rm c}+\overline{\Xi}^{0}_{\rm c}}_\text{raw}/2} 
	= 1- \frac{({\rm Acc}\times\varepsilon)_{\textnormal{feed-down}}\cdot\Delta y\cdot\Delta p_\mathrm{T} \cdot{\rm BR}\cdot \mathcal{L}_{\rm int}}{N^{\Xi^{0}_{\rm c}+\overline{\Xi}^{0}_{\rm c}}_\text{raw}/2}\times \left(\frac{\text{d}^2\sigma}{\text{d}p_\mathrm{T}\text{d}y}\right)_{\Xi^0_{\rm c}\ \textnormal {feed-down}},
	\label{fprompt}
\end{equation}

where $N^{\Xi^{0}_{\rm c}+\overline{\Xi}^{0}_{\rm c}}_\text{raw}/2 $ is the raw yield divided by a factor of two to account for particles and antiparticles, $({\rm Acc}\times\varepsilon)_\textnormal {feed-down}$ is the product of detector acceptance and efficiency for feed-down $\Xi^0_{\rm c}$ baryons and $\left(\frac{\text{d}^2\sigma}{\text{d}p_\mathrm{T}\text{d}y}\right)_{ \Xi^0_{\rm c}\textnormal{ feed-down}}$ is the \pt-differential cross section of feed-down $\Xi^0_{\rm c}$ baryon production. 
The production cross section of $\Xi^0_{\rm c}$ from beauty-baryon decays is not known, hence a strategy based on the estimation made in Ref.~\cite{Acharya:2020uqi} for the cross section of feed-down $\Lambda^+_{\rm c}$ is adopted. The production cross section of $\Lambda^+_{\rm c}$ from $\Lambda^0_{\rm b}$-baryon decays is calculated using the b-quark \pt-differential cross section from FONLL calculations, multiplied by the fraction of beauty quarks that fragment into $\Lambda^0_{\rm b}$. The latter is derived from the LHCb measurement of beauty fragmentation fractions in pp collisions at $\sqrt{s}$ = 13 TeV~\cite{Aaij:2019pqz}, taking into account its \pt dependence.
The $\Lambda^0_{\rm b} \rightarrow \Lambda^+_{\rm c} + {\rm X}$ decay kinematics is modeled using PYTHIA~8.243 simulations~\cite{PYTHIA8}.  The cross section of $\Lambda^+_{\rm c}$ from $\Lambda^0_{\rm b}$-baryon decays is scaled by the fraction of $\Xi^-_{\rm b}$ decaying in a final state with a $\Xi^0_{\rm c}$ over the fraction of $\Lambda^0_{\rm b}$ decaying to $\Lambda^+_{\rm c}$, which are taken to be 50\% and 82\%, respectively, from the PYTHIA~8.243 generator~\cite{PYTHIA8}.
The cross section of $\Xi^0_{\rm c}$ from beauty feed-down is then calculated from the cross section of $\Lambda^+_{\rm c}$ originating from $\Lambda_{\rm b}^{0}$ decays, which is scaled by the ratio of the measured \pt-differential yields of inclusive $\Xi^0_{\rm c}$ and prompt $\Lambda^+_{\rm c}$ baryons.
This procedure relies on the assumptions that the \pt shape of the cross sections of feed-down $\Lambda^+_{\rm c}$ and $\Xi^0_{\rm c}$ is the same and that the ratio $\Xi^0_{\rm c}/\Lambda^+_{\rm c}$ is the same for inclusive and feed-down baryons, along with the consideration that the inclusive $\Lambda^+_{\rm c}$-baryon yield is dominated by the prompt production, based on the $f_{\rm prompt}$ values close to unity reported in Ref.~\cite{Acharya:2020lrg}. The value of $f_{\rm prompt}$ as a function of \pt is shown in the right panel of Fig.~\ref{fig4}.

\subsection{Systematic uncertainties} 

The systematic uncertainty on the $\Xi^0_{\rm c}$ production cross section has different contributions, which are summarised in Table~\ref{tab:systotal} for three representative \pt intervals, namely 2~$<~\pt~<$~3~GeV/$c$, 4~$< \pt <$~5~GeV/$c$, and 6~$< \pt <$~8~GeV/$c$, and discussed in the following.
The overall systematic uncertainty is calculated summing in quadrature the different contributions, which are assumed to be uncorrelated among each other. 

\begin{table}[tb]
\caption{ Contributions to the systematic uncertainty of the $\Xi^{0}_{\rm{c}}$ cross section for the \pt intervals 2~$<~\pt~<$~3~GeV/$c$, 4~$< \pt <$~5~GeV/$c$, and 6~$< \pt <$~8~GeV/$c$.}
	\begin{center}
	\renewcommand\arraystretch{1.2}
	\begin{tabular}{lcccc}
	\hline
	\hline
	  \pt (GeV/$c$) & 2--3 & 4--5 & 6--8\\
	  	\hline
	  ITS-TPC matching &  2\% & 2\% & 2\%  \\
	  Electron track selection &  2\% & 2\%& 2\% \\
	   $\Xi^{\pm}$-daughter track selection & 4\% &4\% & 4\%\\
	  Electron identification &4\% &   7\%  &5\% \\
	  $\Xi^{\pm}$ topological selection & 6\% &6\% & 6\%\\
	  e$\Xi$-pair selection &  3\% & 3\% &3\% \\
	  Bayesian-unfolding iterations & 5\%&9\% & 2\% \\
	  Unfolding method & 5\%&6\% & 4\% \\
	  Response-matrix \pt range and binning	&  6\%  & -- & -- \\
	   $\Xi_{\rm b}$ oversubtraction &  1\% & 1\% & 1\%  \\
	  Generated \pt shape & 2\% &2\% &2\% \\
	  Sensitivity to rapidity interval & 4\% &4\% &4\% \\
	  Feed-down subtraction &  $^{+2}_{-2}$\% & $^{+3}_{-3}$\% & $^{+5}_{-5}$\% \\
	        \hline
      \hline
	 {Total systematic uncertainty }&  $^{+14}_{-14}$\% & $^{+16}_{-16}$\%& $^{+13}_{-13}$\% \\
	\hline
      Branching ratio & & 28.4\% & \\
      Luminosity & & 2.1\% & \\
	\hline
    \hline
	\end{tabular}
	\end{center}
	\label{tab:systotal}
\end{table}

The systematic uncertainty on the tracking efficiency is estimated by comparing the probability of prolonging a track from the TPC to the ITS (“matching efficiency”) in data and simulation, and by varying the track-selection criteria in the analysis.
The uncertainty on the matching efficiency affects only the electron track, and not the tracks of the $\Xi^{-}$ decay particles, for which the prolongation to ITS is not required.  It is defined as the relative difference in the ITS-TPC matching efficiency  between simulation and data. The uncertainty, which slightly depends on the track \pt, is propagated from the electron track to the $\Xi^0_{\rm c}$ taking into account the decay kinematics and is 2\% independent of $\Xi^0_{\rm c}$ \pt.
The second contribution to the track reconstruction uncertainty is estimated by repeating the analysis varying the TPC track selection criteria separately for the electron track and for the $\Xi^{-}$ daughter tracks. The uncertainty is obtained from the root mean square (RMS) of the $\Xi^0_{\rm c}$ cross section values obtained with the different track selection criteria and is 2\% for the electron track and 4\% for the $\Xi^{-}$ daughters independent of $\Xi^0_{\rm c}$ \pt.

Systematic uncertainties can arise from discrepancies in the particle-identification efficiency between simulation and data. The analysis is repeated by varying the selection criteria applied to identify the electron candidate tracks. The systematic uncertainty ranges from 4\% to 7\% depending on the $\Xi^0_{\rm c}$ \pt.

The systematic uncertainty of the efficiency correction for the $\Xi^{-}$ topological selection is 6\% and it is estimated from the RMS of the distribution of the $\Xi^0_{\rm c}$ corrected yields, when the $\Xi^{-}$ topological selection criteria are varied relative to the default measurement.

The uncertainty of the ${\rm e}^+\Xi^-$-pair selection efficiency is estimated by varying the selection criteria of the opening angle and the invariant mass of the pair. A 3\% uncertainty is assigned, independent of $\Xi^0_{\rm c}$ \pt.

The systematic uncertainty of the correction for the missing neutrino momentum is studied testing the stability of the results when varying the unfolding procedure. As a first test, the number of iterations in the Bayesian unfolding procedure is varied. The contribution ranges from 5\% (9\%) at low (intermediate) \pt to 2\% in the highest \pt interval of the measurement.
The second contribution arises from the variation of the unfolding method. The Singular Value Decomposition (SVD) method~\cite{Hocker:1995kb} is used and a \mbox{\pt-dependent} systematic uncertainty  between 4\% and 7\% is assigned based on the difference with respect to the Bayesian method. The last contribution is related to the \pt range and the binning of the response matrix used in the unfolding. 
Systematic uncertainties of 6\% and 4\% are assigned in the
intervals \mbox{$2<\pt<3$ GeV/$c$ }and $3<\pt<4$ GeV/$c$, respectively. At higher \pt, this contribution is negligible.
For these three contributions, the systematic uncertainty is defined as the RMS of the yield values obtained after the unfolding.

The systematic uncertainty due to the subtraction of the $\Xi^{0,-}_{\rm b}$ contribution to the WS pairs is estimated by varying the $\Xi^{0,-}_{\rm b}$ yield and momentum distribution based on the uncertainties of the $\Lambda_{\rm b}^{0}$ \pt-differential cross section in pp collisions~\cite{Chatrchyan:2012xg}.
The assigned systematic uncertainty is 1\%, independent of $\Xi^0_{\rm c}$ $\pt$.

The systematic uncertainty due to the uncertainty of the generated $\Xi^0_{\rm c}$ \pt shape used in the determination of the efficiency is estimated by using the shape from the PYTHIA~8 generator instead of the one from the iterative procedure and is found to be 2\%, independent of \pt. An additional source of uncertainty originates from  possible differences between the $\Xi^0_{\rm c}$-rapidity distributions in data and in the simulation, which affect the measured cross section because the  (Acc~$\times~\varepsilon$) depends on the $\Xi^0_{\rm c}$ rapidity. The systematic uncertainty is estimated to be 4\% by comparing the cross section values obtained  using the values of 
(Acc~$\times~\varepsilon$) and $\Delta y$ obtained considering the generated $\Xi^0_{\rm c}$ baryons in different rapidity intervals (from $|y|<$ 0.5 to $|y|<$ 0.8). 

The systematic uncertainty due to the subtraction of the feed-down from beauty-hadron decays is estimated by considering the uncertainty on the FONLL predictions and by varying the assumption on the ratio $\Xi^0_{\rm c}/\Lambda^+_{\rm c}$ in the $f_{\rm prompt}$ calculation. The FONLL uncertainty is calculated by varying the b-quark mass and the factorisation and renormalisation scales as prescribed in Ref.~\cite{Cacciari:2012ny}. The ratio of inclusive $\Xi^0_{\rm c}$ over prompt $\Lambda^+_{\rm c}$ yield, used to multiply the feed-down $\Xi^{0}_{\rm c}$ cross section, is scaled up by a factor of 2 to account for possible differences between the $\Xi^0_{\rm c}/\Lambda^+_{\rm c}$ and $\Xi_{\rm b}^{0,-}/\Lambda_{\rm b}^{0}$ ratios, and scaled down in order to cover the $\Xi^{0,-}_{\rm b}/\Lambda_{\rm b}^{0}$ value of about 0.12 measured at forward rapidity by the LHCb collaboration~\cite{Aaij:2019ezy}. The uncertainty ranges between 2\% and 5\% depending on the \pt interval. An alternative method for the estimation of the $f_{\rm prompt}$ factor, which consists in the usage of the prompt and feed-down $\Xi^0_{\rm c}$ yields generated with PYTHIA~8 colour reconnection (CR) Mode 2~\cite{Christiansen:2015yqa}, was tested and the obtained results are compatible with the method described above and therefore no systematic uncertainty from this additional method is considered.

All the different sources of systematic uncertainty are considered correlated among the different \pt intervals except the systematic uncertainties due to the unfolding and the pair selection. The \mbox{\pt-differential} cross section has an
additional global normalisation uncertainty due to the uncertainties of the integrated luminosity~\cite{ALICE-PUBLIC-2018-014} and the branching ratio. These contributions are not summed in quadrature with the other sources of uncertainty in Fig.~\ref{fig5} and ~\ref{fig6}.

\section{Results}  \label{sec4}

\begin{figure}[ht!]
\centering
\includegraphics[width=.51\textwidth]{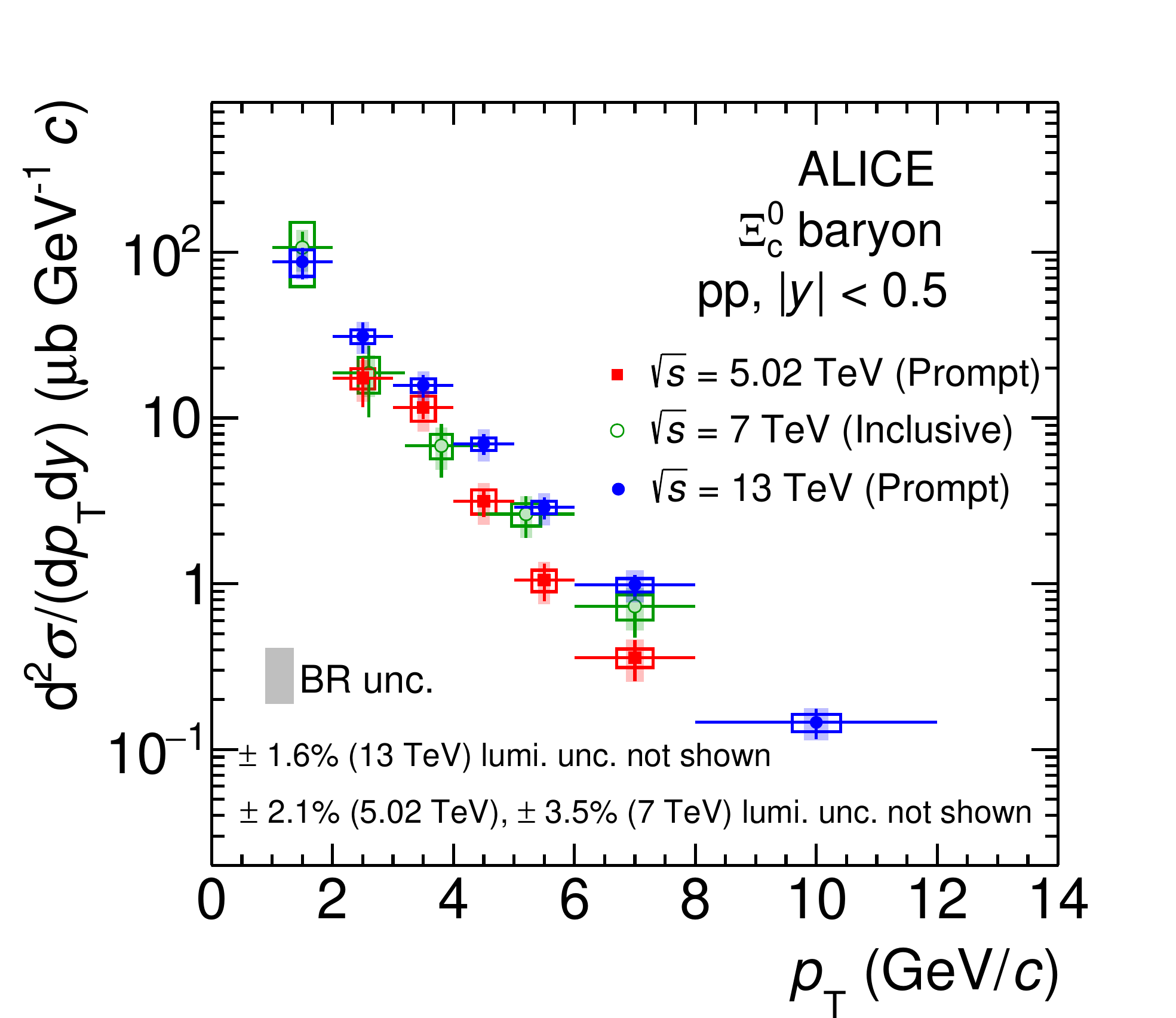}
\hspace{-0.2cm}
\includegraphics[width=.475\textwidth]{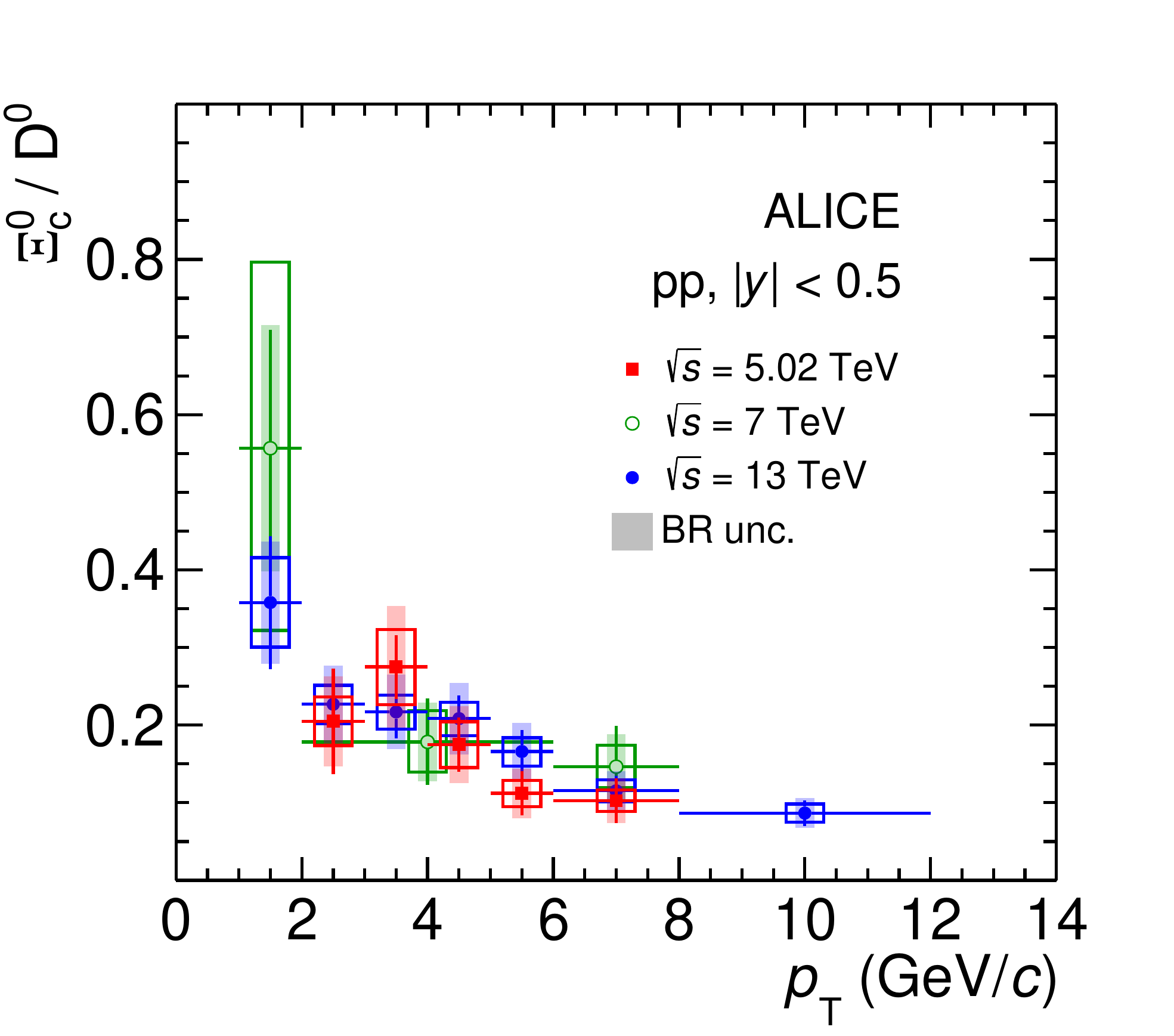}
\caption{Left panel: \pt-differential production cross sections of prompt $\Xi^0_{\rm c}$ baryons in pp collisions at $\sqrt{s}$~=~5.02~TeV and 13 TeV~\cite{xic13tev} and of inclusive $\Xi^0_{\rm c}$ baryons in pp collisions at $\sqrt{s}$~=~7~TeV ~\cite{Acharya:2017lwf} with updated decay BR as discussed in the text. The uncertainty of the BR of the cross sections of prompt $\Xi^0_{\rm c}$ baryons in pp collisions at $\sqrt{s}$~=~13 TeV is lower because it consists in the combination of two different decay channels ($\Xi^0_{\rm c} \rightarrow {\rm e}^{+}\Xi^{-}\nu_{\rm e}$ and $\Xi^0_{\rm c} \rightarrow {\pi}^{+}\Xi^{-}$)~\cite{xic13tev}.
Right panel: $\Xi^0_{\rm c}$/${\rm D^0}$ ratio measured in pp collisions at $\sqrt{s}$~=~5.02~TeV, compared with the measurements at $\sqrt{s}$~=~7~TeV~\cite{Acharya:2017lwf} and $\sqrt{s}$~=~13~TeV~\cite{xic13tev}. The uncertainty of the BR of ${\rm D^0}$ and $\Xi^0_{\rm c}$ are shown as shaded boxes. 
}
\label{fig5}
\end{figure}

The \pt-differential cross section of prompt $\Xi^0_{\rm c}$-baryon production in pp collisions at $\sqrt{s}$ = 5.02 TeV, measured in the rapidity interval $|y|~<$~0.5 and \pt range 2~$< \pt <$~8~GeV/$c$, is shown in the left panel of Fig.~\ref{fig5}. It is compared  with the previously published measurements of inclusive $\Xi^0_{\rm c}$-baryon production in pp collisions at $\sqrt{s}$~=~7~TeV~\cite{Acharya:2017lwf}, updated with the BR value from Ref.~\cite{xic13tev}, and of  prompt $\Xi^0_{\rm c}$-baryon production at $\sqrt{s}$~=~13~TeV~\cite{xic13tev}, which is measured as the average of two decay channels ($\rm \Xi_c^0\rightarrow \Xi^-e^+\nu_e$ and $\rm \Xi_c^0\rightarrow \Xi^{-}\pi^+$). 
The prompt fraction in the  $\Xi^0_{\rm c}$-baryon yield is close to unity (see right panel of Fig.~\ref{fig4}), hence the comparison of the inclusive $\Xi^0_{\rm c}$ cross section measured at $\sqrt{s}$~=~7 TeV with the prompt ones at $\sqrt{s}$~=~5 and 13 TeV provides a meaningful insight into the $\sqrt{s}$ dependence of the production cross section.  
The vertical bars and empty boxes represent the statistical and systematic uncertainties. The systematic uncertainties of the BR are shown as shaded boxes. The uncertainty of the integrated luminosity is not included in the boxes. The data points are positioned at the centre of the \pt intervals. As expected, a smaller $\Xi^0_{\rm c}$ production cross section is measured at lower collision energies. 
The difference between the cross sections at different $\sqrt{s}$ values increases with increasing \pt, indicating a hardening of the \pt-differential spectrum with increasing collision energy.
This behaviour is consistent with that observed for the D-meson and $\Lambda^+_{\rm c}$-baryon cross sections at $\sqrt{s}$ = 5.02, 7 and 13 TeV~\cite{Acharya:2019mgn,Acharya:2017jgo,Acharya:2017kfy,Acharya:2020lrg,SigmacLambdac},  and with the expectations from pQCD calculations~\cite{Cacciari:1998it, Cacciari:2012ny}.
The visible cross section is computed by integrating the \pt-differential cross section in the \pt interval of the measurement.

\begin{equation}
\frac{\text{d}\sigma^{\Xi^0_{\rm c}}_{\text{pp, 5.02 TeV}}}{\text{d}y}{\bigg|}^{(2 < \pt < 8~\text{GeV}/c)}_{|y| < 0.5} = 33.9~\pm~6.0~\text{(stat.)}~\pm~10.6~\text{(syst.)}~\pm~0.7~\text{(lumi.)}~\mu{\rm b}.
	\label{visiblexsec}
\end{equation}

The BR uncertainty is included in the systematic uncertainty.
 
In the right panel of Fig.~\ref{fig5} the $\Xi^0_{\rm c}/{\rm D^0}$ cross section ratio measured in pp collisions at $\sqrt{s}$~=~5.02~TeV as a function of \pt is shown and compared with the same baryon-to-meson ratio measured at $\sqrt{s}$~=~7~\cite{Acharya:2017lwf} and 13~TeV~\cite{xic13tev}.
The prompt ${\rm D^0}$ cross section is reported in Ref.~\cite{Acharya:2021cqv} in finer \pt intervals than those used in the prompt $\Xi^0_{\rm c}$ analysis and is thus rebinned to match the \pt intervals of the $\Xi^0_{\rm c}$ measurement. When merging the ${\rm D^0}$ cross section in different \pt intervals, the systematic uncertainties are propagated considering the yield extraction uncertainty as fully uncorrelated and all the other sources as fully correlated among the \pt intervals. The systematic uncertainty on the $\Xi^0_{\rm c}/{\rm D^0}$ ratio is calculated assuming all the uncertainties of the $\Xi^0_{\rm c}$ and ${\rm D^0}$ cross sections as uncorrelated, except for the tracking and feed-down systematic uncertainties, which partially cancel in the ratio. The uncertainty of the luminosity fully cancels in the baryon-to-meson ratio.
The $\Xi^0_{\rm c}/{\rm D^0}$ ratios at the three centre-of-mass energies are consistent with each other within uncertainties. At low \pt, the ratio is about 0.2 and it decreases with increasing \pt, reaching a value of about 0.1 for $\pt > 6$ GeV/$c$. 
The $\Xi^0_{\rm c}/{\rm D^0}$ ratio in pp collisions at $\sqrt{s}$ = 5.02 TeV integrated in 2~$< \pt <$~8~GeV/$c$ is $0.21 \pm 0.04~{\rm (stat.)}~\pm 0.07~{\rm (syst.)}$, which is calculated as the ratio of the integrated cross sections of $\rm \Xi_c^0$ and $\rm D^0$ in the considered \pt interval. 

\subsection{Comparison with model calculations} 

\begin{figure}[b!]
\centering
\includegraphics[width=0.49\textwidth]{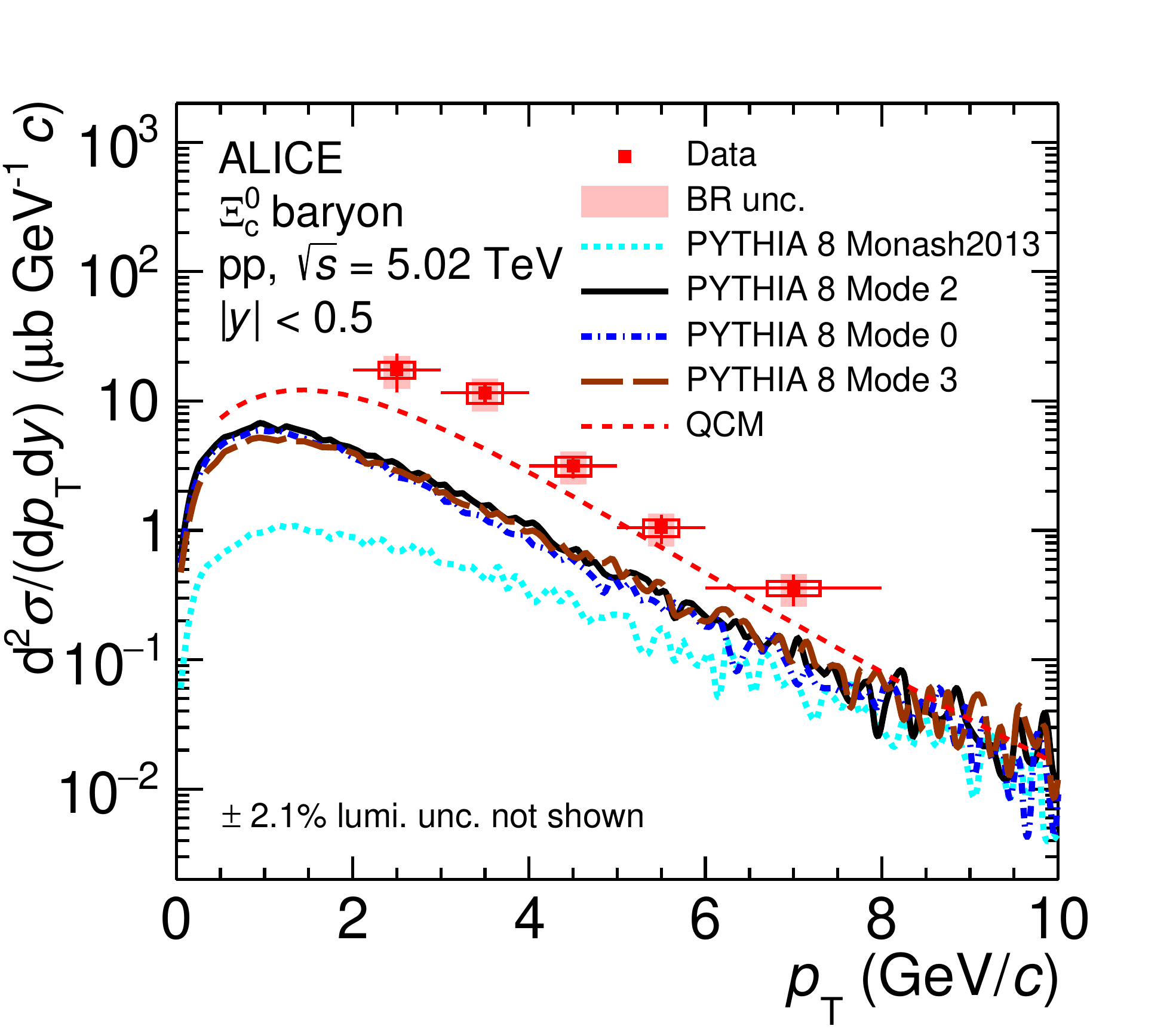}
\includegraphics[width=0.49\textwidth]{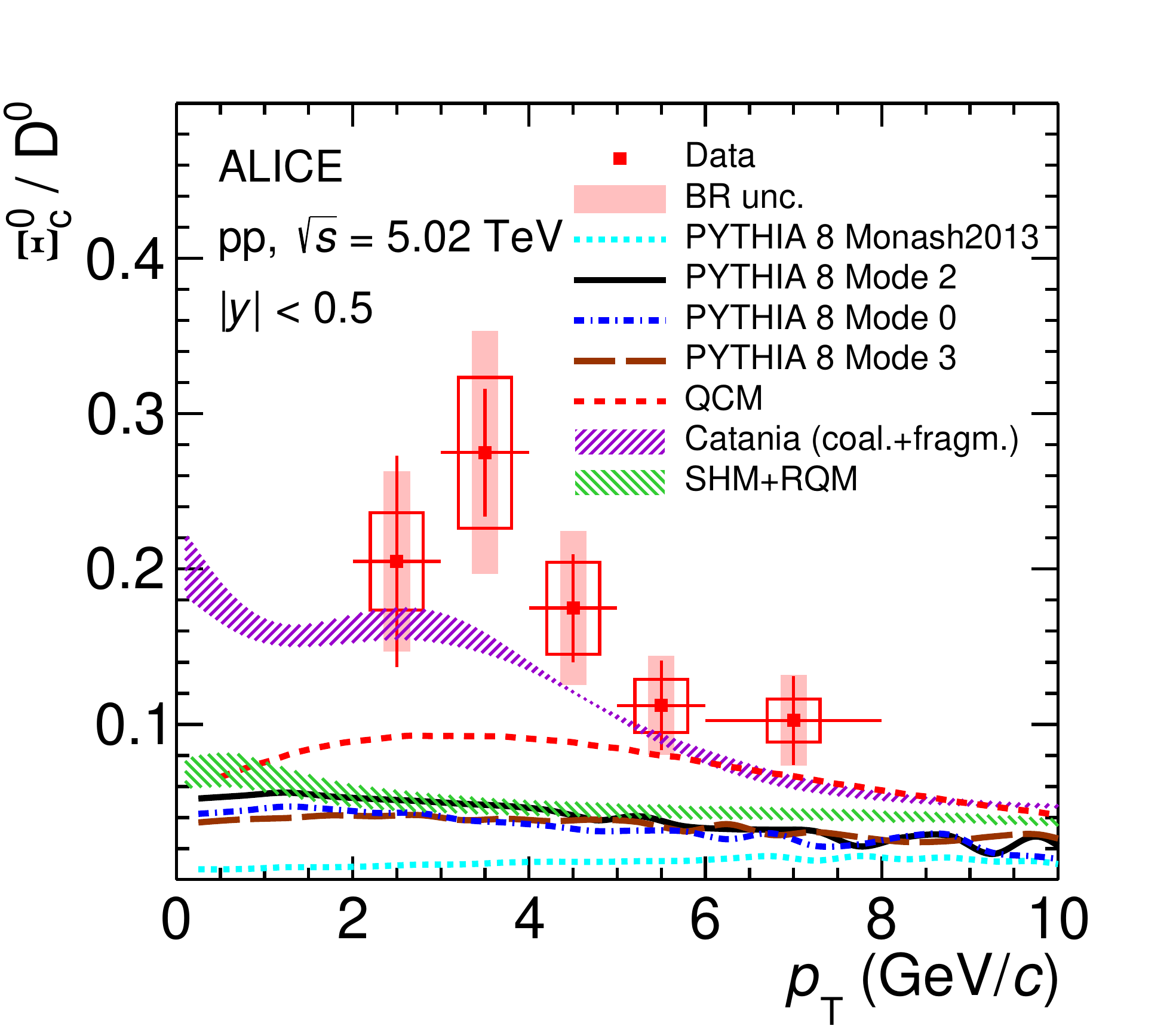}
\caption{Left panel: \pt-differential production cross section of prompt $\Xi^0_{\rm c}$ baryons  in pp collisions at $\sqrt{s}$~=~5.02~TeV compared with model calculations~\cite{Skands:2014pea,Christiansen:2015yqa,Song:2018tpv}. Right panel:  $\Xi^0_{\rm c}/{\rm D^0}$ ratio as a function of \pt measured in pp collisions at $\sqrt{s}$~=~5.02~TeV compared with model calculations~\cite{Skands:2014pea,Christiansen:2015yqa,He:2019tik,Song:2018tpv,Minissale:2020bif} (see text for details).}
\label{fig6}
\end{figure}

The left panel of Fig.~\ref{fig6} shows the comparison of the \pt-differential production cross section of $\Xi^0_{\rm c}$ baryons with predictions from different tunes of the PYTHIA~8.243 generator, including the Monash tune~\cite{Skands:2014pea}, and tunes that implement CR beyond the leading-colour approximation~\cite{Christiansen:2015yqa}. 
In the PYTHIA~8 simulations, all soft QCD processes are enabled. In the Monash tune, the parameters governing the heavy-quark fragmentation are tuned on measurements in ${\rm e^+e^-}$ collisions. The Monash tune significantly underestimates the $\Xi^0_{\rm c}$-baryon production cross section by a factor of about 23 in the lowest \pt interval of the measurement and around a factor 5 in the highest \pt interval.
This prodives an additional information on the non-universality of charm fragmentation that was reported in Refs.~\cite{Acharya:2020uqi,xic13tev,SigmacLambdac} based on the different baryon-to-meson ratios in ${\rm e^+e^-}$ and pp collisions and on the consideration that event generators tuned on ${\rm e^+e^-}$ data do not describe the baryon cross sections measured in pp collisions at LHC energies.
The CR tunes introduce new colour reconnection topologies, including ``junctions'', which favour baryon formation. The three considered tunes (Mode 0, 2, and 3) apply different constraints on the allowed reconnection, taking into account causal connection of dipoles involved in a reconnection and time dilation effects caused by relative boosts between string pieces. It is noted that Mode 2 is recommended in Ref.~\cite{Christiansen:2015yqa} as the standard tune, and contains the strictest constraints on the allowed reconnection. 
The three CR modes yield similar $\Xi^0_{\rm c}$ \pt-differential cross sections, and predict a significantly larger $\Xi^0_{\rm c}$ production cross section with respect to the Monash tune. However, for all three CR modes, the measured $\Xi^0_{\rm c}$ production cross section is underestimated by a factor of about 5--6 for $2<\pt<3$ GeV/$c$, and by a factor of about 3--4 for $\pt > 6$~GeV/$c$, depending on the CR mode. 

The production cross section of the $\Xi^0_{\rm c}$ baryon is also compared with a model using a coalescence approach in hadronic collisions in the framework of QCM~\cite{Gou:2017foe,Song:2018tpv}, in which quarks with equal velocity are combined into hadrons.
A free parameter, $ R^{(\rm c)}_{\rm B/M}$, characterises the relative production of single-charm baryons to single-charm mesons and it is set to 0.425, which is tuned to reproduce the $\Lambda^+_{\rm c}/{\rm D^0}$ ratio measured by ALICE in pp collisions at $\sqrt{s}$ = 7 TeV~\cite{Li:2017zuj}. The relative abundances of the different charm-baryon species are determined by thermal weights. The QCM model is closer to the data as compared to PYTHIA~8 with CR tunes, however it underpredicts the measured cross section by a factor 2--3 for $\pt < 4$~GeV/$c$. 

The measured $\Xi^0_{\rm c}/{\rm D^0}$ ratio is compared in the right panel of Fig.~\ref{fig6} with the different tunes of the PYTHIA~8 event generator previously described. 
All PYTHIA~8 tunes underestimate the measured \pt-differential $\Xi^0_{\rm c}/{\rm D^0}$ ratio. The Monash tune significantly underestimates the data by a factor of about 21--24 in the low \pt region and by a factor of about 7 in the highest \pt interval, as also observed for the $\Lambda^+_{\rm c}/{\rm D^0}$ ratio~\cite{Acharya:2020uqi}. All three CR modes yield a similar magnitude and shape of the $\Xi^0_{\rm c}/{\rm D^0}$ ratio, and despite predicting a larger baryon-to-meson ratio with respect to the Monash tune,  they still underestimate the measured $\Xi^0_{\rm c}/{\rm D^0}$ ratio by a factor of about 4--5 at low $\pt$. The models with CR tunes describe better the $\Lambda^+_{\rm c}/{\rm D^0}$ and the $\Sigma^{0,+,++}_{\rm c}/{\rm D^0}$ ratios than the $\Xi^0_{\rm c}/{\rm D^0}$ one~\cite{Acharya:2021cqv,Acharya:2020uqi,xic13tev,SigmacLambdac}, which involves a charm-strange baryon.

The measured $\Xi^0_{\rm c}/{\rm D^0}$ ratio is also compared with a SHM calculation~\cite{He:2019tik} in which additional excited charm-baryon states not yet observed are included. The additional states are added based on the relativistic quark model (RQM)~\cite{Ebert:2011kk} and lattice QCD calculations~\cite{Briceno:2012wt}. Charm- and strange-quark fugacity factors are used in the model to account for the suppression of quarks heavier than u and d in elementary collisions.
The uncertainty band in the model is obtained by varying the assumption of the branching ratios of excited charm-baryon states decaying to the ground state $\Xi^{0,+}_{\rm c}$, where an exact isospin symmetry between $\Xi^{+}_{\rm c}$ and $\Xi^{0}_{\rm c}$ is assumed.
This model, which was observed  to describe the $\Lambda^+_{\rm c}/{\rm D^0}$ ratio~\cite{Acharya:2020uqi}, underestimates the measured $\Xi^0_{\rm c}/{\rm D^0}$ ratio by the same amount as PYTHIA~8 with CR tunes.

The QCM model~\cite{Song:2018tpv} underpredicts the $\Xi^0_{\rm c}/{\rm D^0}$ ratio by the same amount as it does for the $\Xi^0_{\rm c}$-baryon production cross section.
The Catania model~\cite{Plumari:2017ntm,Minissale:2020bif} implements charm-quark hadronisation via both coalescence and fragmentation. In the model a blast wave parametrisation~\cite{Retiere:2003kf} for light quarks at the hadronisation time with the inclusion of a contribution from mini-jets is considered, while for charm quarks the spectra from FONLL calculations are used. The coalescence process of heavy quarks with light quarks, which is modelled using the Wigner function formalism, is tuned to have all charm quarks hadronising via coalescence at $\pt \simeq 0$. At finite \pt,  charm quarks not undergoing coalescence are hadronised via an independent fragmentation. The Catania model describes the $\Xi^0_{\rm c}/{\rm D^0}$ ratio in the full \pt interval of the measurement.

This new $\Xi^{0}_{\rm c}$ measurement therefore provides important constraints to models of charm quark hadronisation in pp collisions, being in particular sensitive to the description of charm-strange baryon production in the colour reconnection approach, and to the possible contribution of coalescence to charm quark hadronisation in pp collisions.

\subsection{Extrapolation down to  \pt = 0 of the $\Xi^0_{\rm c}$ cross section and the $\Xi^0_{\rm c}/{\rm D^0}$ ratio} 

The \pt-integrated production cross section of prompt $\Xi^0_{\rm c}$ baryons at midrapidity is obtained by extrapolating the visible cross section,
reported in Eq.~\ref{visiblexsec}, to the full \pt range. The PYTHIA~8 generator with CR Mode 2 is used to calculate the central value of the extrapolation factor following what was done for the $\Lambda^+_{\rm c}$ baryon~\cite{Acharya:2020uqi}. 
This prediction was chosen because the PYTHIA 8 generator with CR Mode 2 describes the \pt shape of the measured cross section of $\Xi^0_{\rm c}$ better than the other models that provide predictions of $\Xi^0_{\rm c}$ production in the full \pt range.
The \pt-differential $\Xi^0_{\rm c}$ cross section values for 0 $< \pt <$ 2 GeV/$c$ and for $\pt > 8$~GeV/$c$ are obtained by multiplying the measured $\Xi^0_{\rm c}$ cross section in 2 $< \pt <$ 8 GeV/$c$ by the ratio of the cross sections obtained with PYTHIA~8 in the full and in the measured \pt range.
The systematic uncertainty is estimated from the difference with respect to the extrapolation factors obtained using all the other available model calculations~\cite{Christiansen:2015yqa,He:2019tik,Song:2018tpv,Minissale:2020bif} except for the Monash tune~\cite{Skands:2014pea}, which fails to reproduce the \pt shape of the $\Xi^0_{\rm c}$-baryon cross section. The extrapolation factor is $2.65^{+0.54}_{-0.44}$.
The resulting \pt-integrated cross section of prompt  $\Xi^0_{\rm c}$-baryon production in pp collisions at $\sqrt{s}$~=~5.02~TeV is

\begin{equation}
\frac{\text{d}\sigma^{\Xi^0_{\rm c}}_{\text{pp, 5.02 TeV}}}{\text{d}y}{\bigg|}_{|y| < 0.5} = 89.8~\pm~16.0~\text{(stat.)}~\pm~28.1~\text{(syst.)}~\pm~1.9~\text{(lumi.)\ }^{+18.2}_{-15.0}~\text{(extrap.)}~\mu {\rm b}.
	\label{totalxsec}
\end{equation}

The \pt-integrated cross section is used to calculate the ratio to the one of the $\rm{D^{0}}$ meson which is measured at the same collision energy~\cite{Acharya:2021cqv}. The \pt-integrated  $\Xi^0_{\rm c}/{\rm D^0}$ ratio is $0.20 \pm 0.04~{\rm (stat.)} ^{+0.08}_{-0.07}~{\rm (syst.)}$. In the baryon-to-meson ratio the tracking, the FONLL contribution to the feed-down, and the luminosity components of the systematic uncertainty are considered as correlated between the $\Xi^0_{\rm c}$ and the ${\rm D^0}$ cross sections, while the other sources are treated as uncorrelated. The extrapolation uncertainty is included in the total systematic uncertainty. 
For an accurate measurement of the $\rm c\overline{c}$ production cross section at midrapidity in pp collisions at the LHC, it is therefore necessary to include the large yield of $\Xi^0_{\rm c}$ baryons.

\section{Summary and conclusions}
\label{sec6}
The measurement of the production of prompt $\rm{\Xi_c^0}$ baryons in pp collisions at $\sqrt{s}$ = 5.02 TeV  at midrapidity (\mbox{$|y| <$ 0.5}) with the ALICE detector at the LHC is reported.
The analysis was performed via the semileptonic decay channel $\Xi^{0}_{\rm c} \rightarrow {\rm e^{+}}\Xi^{-}\nu_{\rm e}$ and its charge conjugate. The \pt-differential cross section was measured in the transverse-momentum interval 2 $< \pt <$ 8 GeV/$c$.

The measured \pt-differential cross section and $\rm{\Xi_c^0}/\rm{D^0}$ ratio were compared with different tunes of the PYTHIA~8 event generator that implement different particle production and hadronisation mechanisms. The predictions from the default PYTHIA~8 tune (Monash 2013) and from CR tunes utilising string formation beyond the leading-colour approximation are systematically lower than the experimental measurement. The PYTHIA 8 simulations with the colour-reconnection mechanism predict an enhanced production of baryons and are closer to the data, as compared to the simulation with the Monash tune.
The \pt-differential $\Xi^0_{\rm c}/{\rm D^0}$ ratio was also compared with the statistical hadronisation model, which underestimates the measured ratio also in the case in which the calculations are performed assuming the existence of a large set of yet-unobserved charm-baryon states. Note that PYTHIA 8 with CR and the statistical hadronisation model with additional baryons describe reasonably well the $\Lambda^+_{\rm c}/{\rm D^0}$ ratio.
The measured $\Xi^0_{\rm c}/{\rm D^0}$ ratio is better described by the Catania model, which implements a possible new scenario for pp collisions at LHC energies allowing low-\pt charm quarks to hadronise also via coalescence in addition to the fragmentation mechanism.

The measurements reported in this article provide an additional information of non-universality of charm fragmentation and set important and stringent constraints on models of charm-quark hadronisation in pp collisions.


\newenvironment{acknowledgement}{\relax}{\relax}
\begin{acknowledgement}
\section*{Acknowledgements}

The ALICE Collaboration would like to thank all its engineers and technicians for their invaluable contributions to the construction of the experiment and the CERN accelerator teams for the outstanding performance of the LHC complex.
The ALICE Collaboration gratefully acknowledges the resources and support provided by all Grid centres and the Worldwide LHC Computing Grid (WLCG) collaboration.
The ALICE Collaboration acknowledges the following funding agencies for their support in building and running the ALICE detector:
A. I. Alikhanyan National Science Laboratory (Yerevan Physics Institute) Foundation (ANSL), State Committee of Science and World Federation of Scientists (WFS), Armenia;
Austrian Academy of Sciences, Austrian Science Fund (FWF): [M 2467-N36] and Nationalstiftung f\"{u}r Forschung, Technologie und Entwicklung, Austria;
Ministry of Communications and High Technologies, National Nuclear Research Center, Azerbaijan;
Conselho Nacional de Desenvolvimento Cient\'{\i}fico e Tecnol\'{o}gico (CNPq), Financiadora de Estudos e Projetos (Finep), Funda\c{c}\~{a}o de Amparo \`{a} Pesquisa do Estado de S\~{a}o Paulo (FAPESP) and Universidade Federal do Rio Grande do Sul (UFRGS), Brazil;
Ministry of Education of China (MOEC) , Ministry of Science \& Technology of China (MSTC) and National Natural Science Foundation of China (NSFC), China;
Ministry of Science and Education and Croatian Science Foundation, Croatia;
Centro de Aplicaciones Tecnol\'{o}gicas y Desarrollo Nuclear (CEADEN), Cubaenerg\'{\i}a, Cuba;
Ministry of Education, Youth and Sports of the Czech Republic, Czech Republic;
The Danish Council for Independent Research | Natural Sciences, the VILLUM FONDEN and Danish National Research Foundation (DNRF), Denmark;
Helsinki Institute of Physics (HIP), Finland;
Commissariat \`{a} l'Energie Atomique (CEA) and Institut National de Physique Nucl\'{e}aire et de Physique des Particules (IN2P3) and Centre National de la Recherche Scientifique (CNRS), France;
Bundesministerium f\"{u}r Bildung und Forschung (BMBF) and GSI Helmholtzzentrum f\"{u}r Schwerionenforschung GmbH, Germany;
General Secretariat for Research and Technology, Ministry of Education, Research and Religions, Greece;
National Research, Development and Innovation Office, Hungary;
Department of Atomic Energy Government of India (DAE), Department of Science and Technology, Government of India (DST), University Grants Commission, Government of India (UGC) and Council of Scientific and Industrial Research (CSIR), India;
Indonesian Institute of Science, Indonesia;
Istituto Nazionale di Fisica Nucleare (INFN), Italy;
Institute for Innovative Science and Technology , Nagasaki Institute of Applied Science (IIST), Japanese Ministry of Education, Culture, Sports, Science and Technology (MEXT) and Japan Society for the Promotion of Science (JSPS) KAKENHI, Japan;
Consejo Nacional de Ciencia (CONACYT) y Tecnolog\'{i}a, through Fondo de Cooperaci\'{o}n Internacional en Ciencia y Tecnolog\'{i}a (FONCICYT) and Direcci\'{o}n General de Asuntos del Personal Academico (DGAPA), Mexico;
Nederlandse Organisatie voor Wetenschappelijk Onderzoek (NWO), Netherlands;
The Research Council of Norway, Norway;
Commission on Science and Technology for Sustainable Development in the South (COMSATS), Pakistan;
Pontificia Universidad Cat\'{o}lica del Per\'{u}, Peru;
Ministry of Education and Science, National Science Centre and WUT ID-UB, Poland;
Korea Institute of Science and Technology Information and National Research Foundation of Korea (NRF), Republic of Korea;
Ministry of Education and Scientific Research, Institute of Atomic Physics and Ministry of Research and Innovation and Institute of Atomic Physics, Romania;
Joint Institute for Nuclear Research (JINR), Ministry of Education and Science of the Russian Federation, National Research Centre Kurchatov Institute, Russian Science Foundation and Russian Foundation for Basic Research, Russia;
Ministry of Education, Science, Research and Sport of the Slovak Republic, Slovakia;
National Research Foundation of South Africa, South Africa;
Swedish Research Council (VR) and Knut \& Alice Wallenberg Foundation (KAW), Sweden;
European Organization for Nuclear Research, Switzerland;
Suranaree University of Technology (SUT), National Science and Technology Development Agency (NSDTA) and Office of the Higher Education Commission under NRU project of Thailand, Thailand;
Turkish Energy, Nuclear and Mineral Research Agency (TENMAK), Turkey;
National Academy of  Sciences of Ukraine, Ukraine;
Science and Technology Facilities Council (STFC), United Kingdom;
National Science Foundation of the United States of America (NSF) and United States Department of Energy, Office of Nuclear Physics (DOE NP), United States of America.
\end{acknowledgement}

\bibliographystyle{utphys}   
\bibliography{bibliography}

\newpage
\appendix

%
%

\section{The ALICE Collaboration}
\label{app:collab}
\small
\begin{flushleft} 

S.~Acharya$^{\rm 143}$, 
D.~Adamov\'{a}$^{\rm 98}$, 
A.~Adler$^{\rm 76}$, 
J.~Adolfsson$^{\rm 83}$, 
G.~Aglieri Rinella$^{\rm 35}$, 
M.~Agnello$^{\rm 31}$, 
N.~Agrawal$^{\rm 55}$, 
Z.~Ahammed$^{\rm 143}$, 
S.~Ahmad$^{\rm 16}$, 
S.U.~Ahn$^{\rm 78}$
% ...

\bigskip

\bigskip 

\textbf{\Large Affiliation Notes}

\bigskip 

% ...
$^{\rm V}$ Also at: Institute of Theoretical Physics, University of Wroclaw, Poland\\

\bigskip

\bigskip 

\textbf{\Large Collaboration Institutes}

\bigskip 

% ...
$^{16}$ Department of Physics, Aligarh Muslim University, Aligarh, India\\
% ...
$^{31}$ Dipartimento DISAT del Politecnico and Sezione INFN, Turin, Italy\\
% ...
$^{35}$ European Organization for Nuclear Research (CERN), Geneva, Switzerland\\
% ...
$^{55}$ INFN, Sezione di Bologna, Bologna, Italy\\
% ...
$^{76}$ Johann-Wolfgang-Goethe Universit\"{a}t Frankfurt Institut f\"{u}r Informatik, Fachbereich Informatik und Mathematik, Frankfurt, Germany\\
% ...
$^{78}$ Korea Institute of Science and Technology Information, Daejeon, Republic of Korea\\
% ...
$^{83}$ Lund University Department of Physics, Division of Particle Physics, Lund, Sweden\\
% ...
$^{98}$ Nuclear Physics Institute of the Czech Academy of Sciences, \v{R}e\v{z} u Prahy, Czech Republic\\
% ...
$^{143}$ Variable Energy Cyclotron Centre, Homi Bhabha National Institute, Kolkata, India\\
% ...

\end{flushleft} 
  
\end{document}